\documentclass{pasj01}
\usepackage[round]{natbib}
\usepackage{url}
\usepackage{color}
\draft

\begin{document}
\SetRunningHead{Y. Matsumoto et al.}{MHD Simulation Package: CANS+}

\title{Magnetohydrodynamic Simulation Code CANS+: Assessments and Applications}

\author{Yosuke \textsc{Matsumoto}\altaffilmark{1,2}, Yuta \textsc{Asahina}\altaffilmark{3}, Yuki \textsc{Kudoh}\altaffilmark{4}, Tomohisa \textsc{Kawashima}\altaffilmark{5,6}, Jin \textsc{Matsumoto}\altaffilmark{7}, Hiroyuki R. \textsc{Takahashi}\altaffilmark{8}, Takashi \textsc{Minoshima}\altaffilmark{9}, Seiji \textsc{Zenitani}\altaffilmark{10}, Takahiro \textsc{Miyoshi}\altaffilmark{11}, and Ryoji \textsc{Matsumoto}\altaffilmark{1}}
\altaffiltext{1}{Department of Physics, Graduate School of Science, Chiba University, 1-33 Yayoi-cho, Inage-ku, Chiba 263-8522}
\altaffiltext{2}{Institute for Global Prominent Research, Chiba University, 1-33 Yayoi-cho, Inage-ku, Chiba 263-8522}
\altaffiltext{3}{Center for Computational Sciences, Tsukuba University, 1-1-1 Tennodai Tsukuba-city, Ibaraki 305-8577}
\altaffiltext{4}{Graduate School of Science and Engineering, Kagoshima University, Korimoto 1-21-24, Kagoshima 890-0065}
\altaffiltext{5}{Center for Computational Astrophysics, National Astronomical Observatory of Japan, 2-21-1 Osawa, Mitaka, Tokyo 181-8588}
\altaffiltext{6}{Division of Theoretical Astronomy, National Astronomical Observatory of Japan, 2-21-1 Osawa, Mitaka, Tokyo 181-8588}
\altaffiltext{7}{Research Institute of Stellar Explosive Phenomena, Fukuoka University, 8-19-1 Nanakuma, Jonan-ku, Fukuoka 814-0180}
\altaffiltext{8}{Engineering Science Laboratory, Chubu University, 1200 Matsumoto-cho, Kasugai, Aichi 487-8501}
\altaffiltext{9}{Department of Mathematical Science and Advanced Technology, Japan Agency for Marine-Earth Science and Technology, 3173-25 Syowa-machi, Kanazawaku,Yokohama, 236-0001}
\altaffiltext{10}{Research Institute for Sustainable Humanosphere, Kyoto University, Gokasho, Uji, Kyoto 611-0011}
\altaffiltext{11}{Graduate School of Science, Hiroshima University, 1-3-1 Kagamiyama, Higashi Hiroshima, Hiroshima 739-8526}

\email{ymatumot@chiba-u.jp}


\KeyWords{Magnetohydrodynamics (MHD) --- Methods: numerical --- Shock waves --- Turbulence} 

\maketitle

\begin{abstract}
We present a new magnetohydrodynamic (MHD) simulation code with the aim of providing accurate numerical solutions to astrophysical phenomena where discontinuities, shock waves, and turbulence are inherently important. The code implements the HLLD approximate Riemann solver, the fifth-order-monotonicity-preserving interpolation (MP5) scheme, and the hyperbolic divergence cleaning method for a magnetic field. This choice of schemes significantly improved numerical accuracy and stability, and saved computational costs in multidimensional problems. Numerical tests of one- and two-dimensional problems showed the advantages of using the high-order scheme by comparing with results from a standard second-order TVD MUSCL scheme. The present code enabled us to explore long-term evolution of a three-dimensional accretion disk around a black hole, in which compressible MHD turbulence caused continuous mass accretion via nonlinear growth of the magneto-rotational instability (MRI). Numerical tests with various computational cell sizes exhibited a convergent picture of the early nonlinear growth of the MRI in a global model, and indicated that the MP5 scheme has more than twice the resolution of the MUSCL scheme in practical applications.
\end{abstract}

\section{Introduction}
In the last decades, computational astrophysics has emerged together with the rapid growth of computational capabilities, enabling us to reveal various aspects of astrophysical phenomena that cannot be provided by observations. Among them, magnetohydrodynamic (MHD) simulations are powerful tools for understanding space and astrophysical phenomena such as solar flares and coronal mass ejections, auroral substorms in the terrestrial magnetosphere, magnetic dynamos in accretion disks and associated jet accelerations. In most cases, the systems are dominated by various MHD discontinuities, shock waves, and turbulence in which the fluid dynamics are strongly coupled with the magnetic field. Such highly nonlinear systems have stimulated the search for numerical algorithms that can solve the MHD equations with both high accuracy and stability.

Early development of MHD simulation codes was based on finite difference schemes with artificial and/or numerical dissipation, such as the modified Lax--Wendroff scheme \citep{Rubin1967,Shibata1983}. The high-order upwind scheme was implemented in the ZEUS code \citep{Stone1992,Hawley1995} and has been used for many applications in astrophysics owing to its simplicity and flexibility. However, besides these early successes in computational astrophysics, overcoming spurious (grid) oscillations in highly stratified, compressible MHD applications has remained a task for further improvement in the numerical schemes \citep[e.g.,][]{Kudoh1999}.

Modern MHD simulation codes have a strategy of accurately capturing shock waves as situations often accompany with supersonic flows in space and astrophysical phenomena. Such shock-capturing schemes are based on the finite volume method in which the time evolution of cell-averaged conservative variables of the MHD equations is calculated from numerical fluxes at the cell surfaces. Thus, the accuracy and the robustness rely on the method of obtaining the numerical flux from the cell-averaged conservative variables.

In the numerical flux calculation, upwind natures are incorporated by solving a Riemann problem at the cell interface initiated with the two neighboring cell states. The so-called Godunov schemes are based on the idea that the numerical flux can be obtained by integrating the conservative variables in the Riemann fan following the conservation laws in space and time. For numerical solutions to the Riemann problem, various approximate Riemann solvers have been developed including the Roe's scheme \citep{Roe1981} and a family of Harten--Lax--van Leer (HLL) solvers \citep{Harten1983,Li2005,Miyoshi2005}. Among them, the HLLD approximate Riemann solver of \citet{Miyoshi2005} is now the standard method in modern MHD codes \citep[see e.g.,][]{Kritsuk2011} as it provides tractable and stable numerical solutions.

Special care must be taken in multidimensional MHD simulations as the solenoidal property of the magnetic field cannot be straightforwardly satisfied, and numerical errors from the divergence-free condition severely affect results, in particular, when solving the conservative forms of the MHD equations numerically. While divergence cleaning methods aim at maintaining the numerical errors within minimal levels by making modifications to the base equations \citep{Brackbill1980,Powell1999,Dedner2002}, the constrained-transport (CT) algorithm \citep{Evans1988} accomplishes the divergence-free condition within a level of machine round-off errors by adopting a special discritization for the magnetic field. Nowadays, most MHD codes employ variants of the CT algorithm while preserving the upwind nature for the induction equation of the magnetic field \citep{Londrillo2004,Gardiner2005,Lee2009,Miyoshi2011,Minoshima2015,Lee2017,Minoshima2019}. For further information, readers may consult comprehensive comparisons of the schemes by \citet{Toth2000}, and also by \citet{Miyoshi2011} with more recent schemes.

Extensions of the scheme to a higher-order accuracy have been accomplished by adopting piecewise high-order polynomials to reconstruct variables' profiles in each cell used for the local Riemann problem at the cell interface, with which the total variation diminishing (TVD) property has been necessarily incorporated by adopting various slope limiters, so that the profiles degrade to the first order around discontinuities. The so-called MUSCL scheme \citep{VanLeer1979} and its extension of the third-order PPM scheme \citep{Colella1984} have been widely implemented for practical uses in public  MHD codes as FLASH \citep{Lee2009,Lee2013}, PLUTO \citep{Mignone2007}, and Athena \citep{Stone2008}. See also \citet{Kritsuk2011} for other MHD codes. Note, however, that the overall spacial accuracy of these CT-based MHD codes has been achieved only up to the second order because the original CT algorithm is based on the second-order discritization. Third-order scheme has been proposed in the framework of the upwind CT (UCT) with additional large computational costs arising from high-order reconstructions and solutions of the Riemann problem at the cell edge \citep{Londrillo2004}. A simple and cost-effective, fifth-order scheme was proposed by adopting the hyperbolic divergence cleaning method \citep{Dedner2002} by \citet{Mignone2010}. We have adopted a similar strategy as described in the present paper.

The Coordinated Astronomical Numerical Software, CANS, was developed for the Japanese astrophysical community by T. Yokoyama at the University of Tokyo and collaborators, and is publicly available with documents at the website \url{http://www-space.eps.s.u-tokyo.ac.jp/~yokoyama/etc/cans/}. The base schemes are the modified Lax--Wendroff scheme and the CIP--MOCCT scheme \citep{Yabe2001,Kudoh1999}, but the Roe--MUSCL scheme is also available. Additional physics modules of heat conduction and radiative cooling, and many application modules are also included. A physical problem can be easily solved by choosing the numerical schemes and visualizing by prebuilt Interactive Data Language (IDL) procedures, from which researchers and graduate students benefit when starting their simulation studies. CANS has been used in space and astrophysical applications since its start in 2002 \citep[e.g.,][]
{Asai2004,Isobe2005,Hanayama2005,Tao2005,Tsubouchi2009,Toriumi2011}.

CANS+ (CANS-plus) was designed based on the CANS philosophy, but has implemented the state-of-the-art numerical algorithms and parallelization. The codes are written in Fortran90/95 and are organized in a modular way as in CANS. Hybrid parallelization is implemented by using the MPI library and OpenMP for effective parallel scaling on modern massively parallel supercomputer systems. Scripts for reading output data are prepared for both Python and IDL languages so users can choose either of these environments for analysis. Sample codes for visualization in Python using the Matplotlib modules and IDL are also available for quick analyses. The code can be downloaded from the CANS+ documentation website \url{http://www.astro.phys.s.chiba-u.ac.jp/cans/doc/}. In this paper, we detail the numerical algorithms adopted in CANS+ in Section 2 and show assessments of the code's capability with various physical problems in Section 3. As an application of CANS+, we present three-dimensional (3D) global simulations of a black hole accretion disk and discuss how the selection of the numerical schemes affects the angular momentum transport and the resulting mass accretion rate during its long-term evolution in Section 4. A summary of the paper and future perspectives are given in Section 5.


\section{CANS+}

\subsection{Basic equations}
CANS+ solves the MHD equations in a normalized, semiconservative form as

\begin{eqnarray}
\frac{\partial \rho}{\partial t} + \nabla \cdot \left( \rho \mbox{\boldmath$v$} \right) &=& 0, \label{eq:mass}\\
\frac{\partial \rho\mbox{\boldmath$v$}}{\partial t} + \nabla \cdot \left( \rho\mbox{\boldmath$vv$} + p_t\mbox{\boldmath$I$} - \mbox{\boldmath$BB$} \right) &=& \rho\boldsymbol{g}, \label{eq:momentum}\\
\frac{\partial \mbox{\boldmath$B$}}{\partial t} + \nabla \cdot \left( \mbox{\boldmath$vB$}-\mbox{\boldmath$Bv$} +\psi \mbox{\boldmath$I$}\right) &=& -\nabla \times \left( \eta \mbox{\boldmath$j$} \right), \label{eq:induction}\\
\frac{\partial e}{\partial t} + \nabla \cdot \left( (e+p_t)\mbox{\boldmath$v$}-\mbox{\boldmath$B$}(\mbox{\boldmath$v$}\cdot\mbox{\boldmath$B$}) \right) &=& -\nabla \cdot \left(\eta \boldsymbol{j}\times\boldsymbol{B}\right)+\rho\boldsymbol{v}\cdot\boldsymbol{g},\label{eq:energy}\\
\frac{\partial \psi}{\partial t} +c_h^2 \nabla \cdot \mbox{\boldmath$B$} &=& -\frac{c_h^2}{c_p^2}\psi,\label{eq:psi} 
\end{eqnarray}
where $\rho$, $\boldsymbol{v}$, $\boldsymbol{g}$, $\boldsymbol{B}$, and $e$ are the mass density, the velocity, the gravitational acceleration, the magnetic field, and the total energy density, respectively, $\boldsymbol{I}$ is a unit tensor 
\begin{equation}
\boldsymbol{I} = \left(
\begin{array}{ccc}
1 & 0 & 0 \\
0 & 1 & 0 \\
0 & 0 & 1 
\end{array}
\right),
\end{equation}
$\eta$ is the resistivity, $\boldsymbol{j}=\nabla\times\boldsymbol{B}$ is the current density, and $p_t$ represents the total (thermal and magnetic) pressure defined as

\begin{equation}
p_t = p + \frac{B^2}{2}.
\end{equation}
The thermal pressure $p$ is given by

\begin{equation}
\label{eq:pressure}
p = \left(\gamma-1\right) \left(e-\frac{1}{2}\rho v^2- \frac{B^2}{2} \right),
\end{equation}
where $\gamma$ is the specific heat ratio. 

The equation (\ref{eq:psi}) for $\psi$ with equation (\ref{eq:induction}) is introduced so that the solenoidal property of the magnetic field

\begin{equation}
\label{eq:divb}
\nabla \cdot \mbox{\boldmath$B$} = 0
\end{equation}
is maintained within minimal errors during time integration (see subsection \ref{divb}). The equations (\ref{eq:mass}) -- (\ref{eq:psi}) complete a set of the MHD equations, which is conventionally referred to as GLM (generalized
Lagrange multiplier)--MHD equations \citep{Dedner2002}.

The GLM--MHD equations then read

\begin{equation}
\label{eq:eqsys}
\frac{\partial \mbox{\boldmath$U$}}{\partial t} + \sum_{s=x,y,z} \frac{\partial \boldsymbol{F}_s}{\partial s} = \mbox{\boldmath$S$}
\end{equation}
where

\begin{equation}
\mbox{\boldmath$U$} = \left(
\begin{array}{ccccc}
\rho & \rho\mbox{\boldmath$v$} & \mbox{\boldmath$B$} & e & \psi
\end{array}
\right)^T
\end{equation}
is a state vector of the conservative variables,
\begin{equation}
\boldsymbol{F}_s = \left(
\begin{array}{c}
\rho v_s \\
\rho v_s v_x + p_T\delta_{sx} - B_s B_x \\
\rho v_s v_y + p_T\delta_{sy} - B_s B_y \\
\rho v_s v_z + p_T\delta_{sz} - B_s B_z \\
v_s B_x - B_s v_x + \psi \delta_{sx}\\
v_s B_y - B_s v_y + \psi \delta_{sy}\\
v_s B_z - B_s v_z + \psi \delta_{sz}\\
\left( e + p_T \right) v_s - B_s \left( \mbox{\boldmath$v$} \cdot \mbox{\boldmath$B$} \right) \\
c_h^2B_s
\end{array}
\right),
\end{equation}
\begin{equation}
\delta_{ij} = \Bigg\{ 
\begin{array}{c}
1\ \ (i = j) \\
0\ \ (i \ne j) \\
\end{array},
\end{equation}
are a flux vector and the Kronecker delta, respectively, and

\begin{equation}
\label{eq:source}
\mbox{\boldmath$S$} = \left(
\begin{array}{cccccc}
0 & \rho \boldsymbol{g} & -\nabla \times (\eta \boldsymbol{j}) & -\nabla \cdot \left(\eta \boldsymbol{j}\times\boldsymbol{B}\right)+\rho\boldsymbol{v}\cdot\boldsymbol{g} & -\frac{c_h^2}{c_p^2}\psi
\end{array}
\right)^T,
\end{equation}
is a source vector. We also define a vector $\boldsymbol{V}$ for the primitive variables 

\begin{equation}
\mbox{\boldmath$V$} = \left(
\begin{array}{ccccc}
\rho & \mbox{\boldmath$v$} & \mbox{\boldmath$B$} & p & \psi
\end{array}
\right)^T.
\end{equation}

Let us consider equation (\ref{eq:eqsys}) with $\boldsymbol{S}=0$ in one dimension (1D) ($s=x$). It can be written in the form

\begin{equation}
\label{eq:1dsyseq}
\frac{\partial \mbox{\boldmath$U$}}{\partial t} + A\frac{\partial \boldsymbol{U}}{\partial x} = \frac{\partial \mbox{\boldmath$V$}}{\partial t} + A_p\frac{\partial \boldsymbol{V}}{\partial x}=0,
\end{equation}
where
\begin{equation}
A = \frac{\partial \boldsymbol{F}_x}{\partial \boldsymbol{U}}
\end{equation}
and
\begin{equation}
A_p = \left(\frac{\partial \boldsymbol{U}}{\partial \boldsymbol{V}}\right)^{-1}A\frac{\partial \boldsymbol{U}}{\partial \boldsymbol{V}}
\end{equation}
are the Jacobian matrix for the conservative and the primitive variables, respectively, and
\begin{equation}
\frac{\partial \boldsymbol{U}}{\partial \boldsymbol{V}} = \left(
\begin{array}{ccccc}
1 & 0 & 0 & 0 & 0 \\
\boldsymbol{v} & \rho & 0 & 0 & 0 \\
0 & 0 & 1 & 0 & 0 \\
\frac{\boldsymbol{v}\cdot\boldsymbol{v}}{2} & \rho \boldsymbol{v}& \boldsymbol{B} & \frac{1}{\gamma-1} & 0\\
0 & 0 & 0 & 0 & 1 \\
\end{array}
\right)
\end{equation}
is the quasilinear conversion matrix, which relates $\delta\boldsymbol{V}$ to $\delta\boldsymbol{U}$. Multiplying equation (\ref{eq:1dsyseq}) by the left eigenvector ($\boldsymbol{L}_p$ with $\boldsymbol{L}_p \boldsymbol{R}_p=\boldsymbol{I}$) of the Jacobian matrix gives the equation for the characteristic variables $\boldsymbol{W}$ as
\begin{equation}
\boldsymbol{L}_p\frac{\partial \mbox{\boldmath$V$}}{\partial t} + \boldsymbol{L}_pA_p\boldsymbol{R}_p\boldsymbol{L}_p\frac{\partial \boldsymbol{V}}{\partial x}=\frac{\partial \mbox{\boldmath$W$}}{\partial t} + \Lambda\frac{\partial \boldsymbol{W}}{\partial x}=0,
\end{equation}
where $\delta\boldsymbol{W}=\boldsymbol{L}_p\delta\boldsymbol{V}$, and $\boldsymbol{L}_pA_p\boldsymbol{R}_p=\Lambda = diag(\lambda_1,\lambda_2,...,\lambda_9)$ is a diagonal matrix containing the eigenvalues. For the GLM--MHD equations, $\lambda_1 \sim \lambda_9$ are
\begin{equation}
\lambda_{1,9} = \mp c_h,\ \ \lambda_{2,8} = v_x \mp c_f, \ \lambda_{3,7} = v_x \mp c_a,\ \lambda_{4,6} = v_x \mp c_s, \ \lambda_{5} = v_x,
\end{equation}
where
\begin{equation}
c_{f,s} = \left(\frac{\gamma p + |\boldsymbol{B}|^2 \pm \sqrt{\left(\gamma p + |\boldsymbol{B}|^2 \right)^2 - 4\gamma pB_x^2}}{2\rho}\right)^{1/2}
\end{equation}
refer to the fast (positive) and slow (negative) magnetosonic speeds, respectively, and
\begin{equation}
c_{a} = \frac{|B_x|}{\sqrt{\rho}}
\end{equation}
is the Alfv\'en speed. In addition to the fast ($\lambda_{2,8}$), the Alfv\'en ($\lambda_{3,7}$), the slow ($\lambda_{4,6}$), and the entropy ($\lambda_5$) waves of the ideal MHD system, the GLM--MHD system introduces two waves with the eigenvalues (phase speeds) of $\lambda_{1,9} = \mp c_h$, which represent omnidirectional propagation of numerical errors of the divergence-free condition (equation (\ref{eq:divb})).

\subsection{Numerical algorithms}
\label{algorithm}
Time evolution of the equation (\ref{eq:eqsys}) for the state vector at a cell $\boldsymbol{U}_{i,j,k}$ is obtained by the finite volume method

\begin{equation}
\label{eq:fv}
\frac{\partial \boldsymbol{\bar U}_{(i,j,k)}}{\partial t} = - \sum_{s=i,j,k} \frac{\mbox{\boldmath$F$}^{*}_{(s+1/2)}-\mbox{\boldmath$F$}^{*}_{(s-1/2)}}{\Delta_s}+\mbox{\boldmath$S$}_{(i,j,k)} \equiv L(\boldsymbol{\bar U}),
\end{equation}
where $\boldsymbol{\bar U}$ is the cell-averaged value of $\boldsymbol{U}$, $\Delta_s$ with $s=i,j,k$ represents cell width in each dimension, $F^{*}_{(s\pm1/2)}$ denotes a numerical flux at cell surfaces, and $L(\boldsymbol{\bar U})$ defines a numerical operator of the finite volume method. Equation (\ref{eq:fv}) is integrated with time by the third-order strong-stability-preserving (SSP) Runge--Kutta (RK) scheme \citep{Suresh1997,Gottlieb1998}:
\begin{eqnarray}
\boldsymbol{u}^{(0)} &=& \boldsymbol{\bar U}^n, \label{eq:rk0}\\
\boldsymbol{u}^{(1)} &=& \boldsymbol{u}^{(0)}+ \Delta t^n L(\boldsymbol{u}^{(0)}), \label{eq:rk1}\\
\boldsymbol{u}^{(2)} &=& \frac{3}{4}\boldsymbol{u}^{(0)}+\frac{1}{4}\left(\boldsymbol{u}^{(1)}+\Delta t^nL(\boldsymbol{u}^{(1)})\right), \label{eq:rk2}\\ 
\boldsymbol{u}^{(3)} &=& \frac{1}{3}\boldsymbol{u}^{(0)}+\frac{2}{3}\left(\boldsymbol{u}^{(2)}+\Delta t^nL(\boldsymbol{u}^{(2)})\right), \label{eq:rk3}\\
\boldsymbol{\bar U}^{n+1} &=& \boldsymbol{u}^{(3)} \label{eq:rk4},
\end{eqnarray}
where the superscript $n$ stands for a number of time steps and $\Delta t^n$ is a corresponding step size. $\Delta t^n$ is determined from the Courant--Friedrichs--Lewy (CFL) condition
\begin{equation}
\label{eq:cfl}
\Delta t = \sigma_c  \min_{i,j,k}\left( \frac{\Delta_i}{\max(|\lambda_{i2}|,|\lambda_{i8}|)}, 
                                              \frac{\Delta_j}{\max(|\lambda_{j2}|,|\lambda_{j8}|)}, 
                                              \frac{\Delta_k}{\max(|\lambda_{k2}|,|\lambda_{k8}|)} \right),
\end{equation}
where $\lambda_{s2,8}$ ($s=i,j,k$) is the eigenvalue for the fast magnetosonic wave calculated for each dimension, and $\sigma_c$ is the CFL number. For the resistive MHD case, i.e., $\eta \neq 0$, $\Delta t$ from equation (\ref{eq:cfl}) is again compared with the one determined from the diffusion number $\sigma_{d}$
\begin{equation}
\Delta t = \min\left( \Delta t, \sigma_d \min_{i,j,k}\left( \frac{\Delta_i^2}{\eta_{(i,j,k)}}, \frac{\Delta_j^2}{\eta_{(i,j,k)}}, \frac{\Delta_k^2}{\eta_{(i,j,k)}} \right) \right).
\end{equation}
In the following numerical experiments, $\sigma_c=\sigma_d=0.3$ were adopted unless otherwise stated.

The numerical operator $L(\boldsymbol{\bar U})$ is evaluated by following procedures.
\begin{enumerate}
\item Conversion from the cell-averaged primitive variables to the characteristic variables locally \citep{Harten1987}. This operation is done by multiplying the primitive variables at cells within a stencil centered at a cell ($i,j,k$) by the left eigenvector calculated at the cell ($i,j,k$). For example, in the $x$-direction,
\begin{equation}
\boldsymbol{W}_{(i+q,j,k)} = \boldsymbol{L}_{p(i,j,k)} \boldsymbol{\bar V}_{(i+q,j,k)}
\end{equation}
with $q$ varying from $-r$ to $+r$ for the ($2r+1$)th-order scheme. Among various eigenvectors, CANS+ adopts the same eigenvectors as those used in the Athena code \citep{Stone2008}. Although this conversion is computationally expensive, separating waves according to the characteristics is necessary for obtaining nonoscillatory profiles by higher-order reconstructions as in the following.
\item Reconstruction of the characteristic variables $\boldsymbol{W}$ within a cell. To obtain high-resolution, non-oscillatory profiles, the fifth-order-monotonicity-preserving (MP5) interpolation scheme \citep{Suresh1997} is used (see subsection \ref{mp5}). In the MP5 scheme, a five-point stencil ($r=2$) is required for the conversion and the reconstruction steps. The interpolated values at both cell surfaces are obtained at once in each dimension. For example, for a cell at ($i,j,k$) in the x-direction, the left-hand, right-state vector $^R \boldsymbol{W}_{(i-1/2,j,k)}$ and the right-hand, left-state vector $^L \boldsymbol{W}_{(i+1/2,j,k)}$ can be obtained for each interpolation procedure.
\item The interpolated values at cell surfaces are then converted to the primitive variables using the local right eigenvector as
\begin{eqnarray}
^R \boldsymbol{V}_{(i-1/2,j,k)} &=& \boldsymbol{R}_{px(i,j,k)}\ ^R \boldsymbol{W}_{(i-1/2,j,k)}, \\
^L \boldsymbol{V}_{(i+1/2,j,k)} &=& \boldsymbol{R}_{px(i,j,k)}\ ^L \boldsymbol{W}_{(i+1/2,j,k)}, \\
^R \boldsymbol{V}_{(i,j-1/2,k)} &=& \boldsymbol{R}_{py(i,j,k)}\ ^R \boldsymbol{W}_{(i,j-1/2,k)}, \\
^L \boldsymbol{V}_{(i,j+1/2,k)} &=& \boldsymbol{R}_{py(i,j,k)}\ ^L \boldsymbol{W}_{(i,j+1/2,k)}, \\
^R \boldsymbol{V}_{(i,j,k-1/2)} &=& \boldsymbol{R}_{pz(i,j,k)}\ ^R \boldsymbol{W}_{(i,j,k-1/2)}, \\
^L \boldsymbol{V}_{(i,j,k+1/2)} &=& \boldsymbol{R}_{pz(i,j,k)}\ ^L \boldsymbol{W}_{(i,j,k+1/2)},
\end{eqnarray}
where $\boldsymbol{R}_{ps}$ with $s=x,y,z$ represents the right eigenvector in each direction.

\item Evaluation of the numerical flux $\boldsymbol{F}^*$ in each dimension by the Godunov-type scheme. That is, the numerical flux at a cell surface $(i+1/2,j,k)$, for example, can be obtained from
\begin{equation}
\label{eq:flux}
\boldsymbol{F}_{i+1/2}^* = \boldsymbol{F}(\boldsymbol{\bar U}_{i}^n)-\frac{1}{\Delta t^n} \int_{x_i}^{x_{i+1/2}} \boldsymbol{U}_{R}\left(\frac{x-x_{i+1/2}}{\Delta t};\boldsymbol{^L V}_{i+1/2},\boldsymbol{^R V}_{i+1/2}\right) dx+\frac{x_{i+1/2}-x_i}{\Delta t^n}\boldsymbol{\bar U}^n_{i},
\end{equation}
where $\boldsymbol{U}_{R}\left(\frac{x-x_{i+1/2}}{\Delta t};\boldsymbol{^L \bar V}_{i+1/2},\boldsymbol{^R V}_{i+1/2}\right)$ is a solution inside the Riemann fan during a time interval of $\Delta t$ with the initial conditions of the left ($\boldsymbol{^L V}_{i+1/2}$) and right ($\boldsymbol{^R V}_{i+1/2}$) states. (Here we have omitted $(j,k)$ in the notations for simplicity.) CANS+ employs the HLLD approximate Riemann solver of \citet{Miyoshi2005}, which gives accurate and robust solutions to Riemann problems. Note, however, that equations for the magnetic field component normal to the cell surface and for $\psi$ are decoupled from the seven remaining equations. The numerical flux values for these variables are easily obtained separately (see subsection \ref{divb}).
\item The source term is evaluated at each Runge--Kutta substage (equations (\ref{eq:rk1}) -- (\ref{eq:rk3})).
\item Conversion from the updated conservative variables to the primitive variables.
\item Return to 1.
\end{enumerate}

\subsection{The MP5 scheme}
\label{mp5}
\begin{figure}
  \begin{center}
  \includegraphics[scale=0.4]{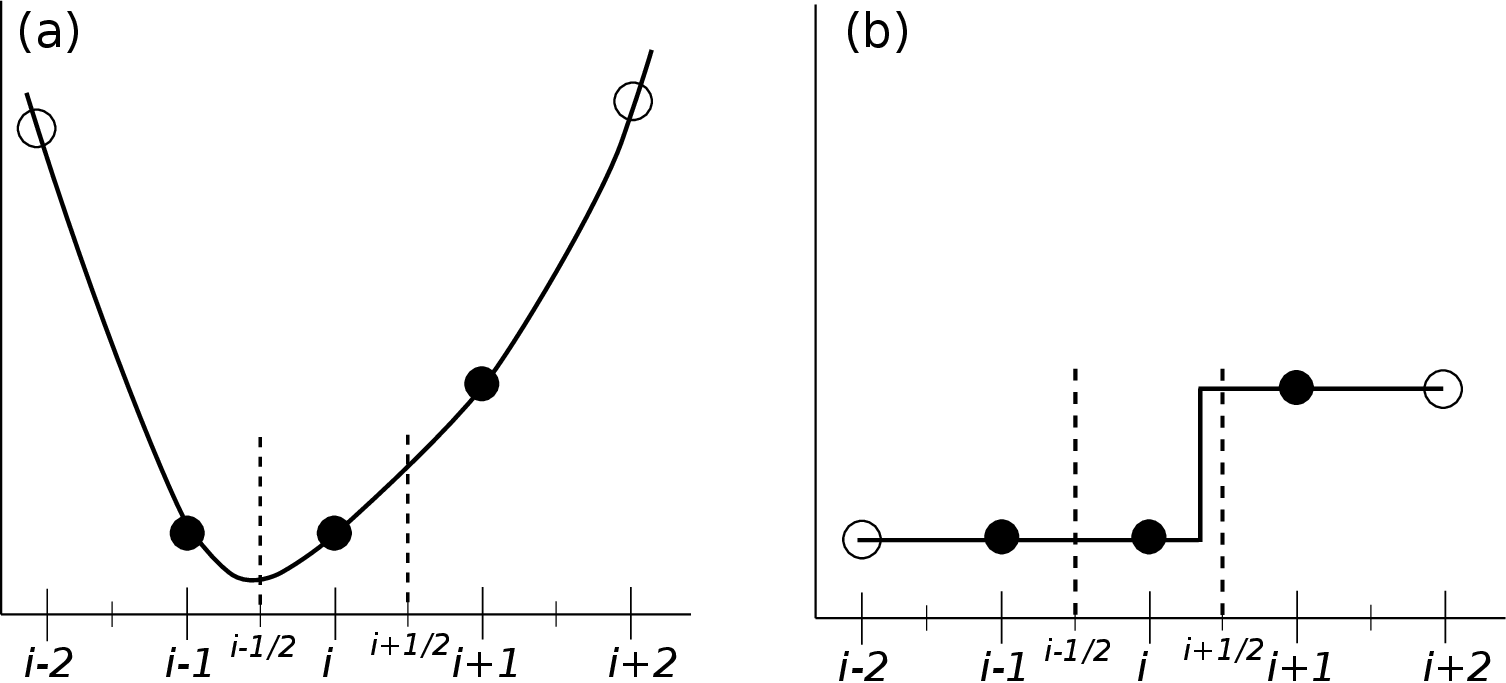}
  \end{center}
  \caption{Profile reconstructions around a cell $i$ using different numbers of stencils (filled and open circles) for (a) parabola and (b) stair-step profiles. Vertical dashed lines indicate cell boundaries where reconstructed data are needed. This figure is adapted from figure 2.1 in \citet{Suresh1997}}.
  \label{fig:mp5}
\end{figure}

The MP5 scheme in the reconstruction step is based on a fourth-degree polynomial for each dimension. The left state of a quantity $f$ at a cell $i+1/2$ is first evaluated by
\begin{equation}
\label{eq:mp5_1}
^L f_{i+1/2} = \frac{2\bar f_{i-2}-13\bar f_{i-1}+47\bar f_i+27\bar f_{i+1}-3\bar f_{i+2}}{60},
\end{equation}
where $\bar f$ is a cell-averaged value of $f$. The right state at $i-1/2$ is also obtained at the same time from its symmetry in each dimension:
\begin{equation}
\label{eq:mp5_2}
^R f_{i-1/2} = \frac{2\bar f_{i+2}-13\bar f_{i+1}+47\bar f_i+27\bar f_{i-1}-3\bar f_{i-2}}{60}.
\end{equation}
The original value results in a fifth-order spatially accurate solution in smooth regions, but an oscillatory profile around discontinuities. The MP5 scheme seeks a profile that degrades to the first order only around discontinuities. For this purpose, the original value is then brought to an interval using the median function
\begin{equation}
\label{eq:median}
^L f_{i+1/2} = {\rm median}(^L f_{i+1/2},f_{min},f_{max}).
\end{equation}
Here the median function returns an intermediate value among three variables in the arguments, so that the cell surface value lies in an interval between $f_{min}$ and $f_{max}$. The basic concept is therefore the same as in standard TVD schemes. The TVD schemes with a three-point stencil, however, cannot distinguish between a discontinuity and a smooth profile around the extremum (figure \ref{fig:mp5}). The interpolated value around the extremum of a sinusoidal profile (figure \ref{fig:mp5}(a)) is then bounded by the neighboring cell values, resulting in strong damping of the waveform (figure \ref{fig:mp5}(b)). Because of this difficulty, the TVD schemes have been applied mainly to problems where strong shock waves dominate. Accuracy and monotonicity preservation of profiles are accomplished in the MP5 scheme by using two additional values at cells (open circles in figure \ref{fig:mp5}) to distinguish between the two. The curvature of a profile can be calculated by using the five-point stencil, and the interval is expanded in cases when the profile is identified as parabola. Analytic consideration of the monotonicity preservation under the SSP--RK time integration for a linear advection equation leads to a CFL number restriction $\sigma_c<0.2$ in the MP5 scheme. In practice, $\sigma_c = 0.3$, which was adopted in the following numerical tests, still gives nonoscillatory results. Readers may consult the original article by \citet{Suresh1997} for detailed derivations of $f_{min}$ and $f_{max}$, and the CFL restriction. Implementation of the MP5 scheme to CANS+ for practical uses is given in Appendix \ref{appendix_crt}.

\subsection{Hyperbolic divergence cleaning method}
\label{divb}
Upwind schemes including the finite volume method always suffer from numerical errors of the solenoidal property of the magnetic field. This occurs because the Lorentz force term in the momentum equation (equation (\ref{eq:momentum})) involves an acceleration term along the magnetic field arising from numerical errors of $\boldsymbol{\nabla}\cdot\boldsymbol{B}$; that is, 
\begin{equation}
 -\nabla \cdot \left( \frac{|\mbox{\boldmath$B$}|^2}{2}\mbox{\boldmath$I$}-\mbox{\boldmath$BB$}\right) = \left(\nabla \times \mbox{\boldmath$B$}\right) \times \mbox{\boldmath$B$} + \mbox{\boldmath$B$}\left(\nabla \cdot \mbox{\boldmath$B$}\right),
\label{eq:divb_origin}
\end{equation}
which finally leads to undesired results and thus inhibits following the long-term evolution of multidimensional problems. To overcome this inherent problem in multidimensional MHD simulations, several approaches have been developed \citep{Toth2000,Miyoshi2011}. Among them, modern MHD simulation codes have adopted strategies to satisfy discretized forms of the divergence-free condition within machine round-off errors based on a family of upwind constrained-transport (CT) algorithms \citep{Londrillo2004,Gardiner2005,Mignone2007,Lee2009,Miyoshi2011,Minoshima2015,Lee2017,Minoshima2019}. In contrast, CANS+ adopts the hyperbolic divergence cleaning method \citep{Dedner2002,Mignone2010}, which transports and damps numerical errors of the divergence-free condition, so that the growth of errors is managed to be within minimal levels. More specifically, multiplying the equation (\ref{eq:psi}) by $\partial/\partial t$ and equation (\ref{eq:induction}) by $\boldsymbol{\nabla}\cdot$, one finds that the scalar potential $\psi$ and $\boldsymbol{\nabla}\cdot\boldsymbol{B}$ obey the telegraph equations
\begin{eqnarray}
& &\frac{\partial^2 \psi}{\partial t^2} + \frac{c_h^2}{c_p^2}\frac{\partial \psi}{\partial t} - c_h^2 \nabla^2\psi=0,\\
& &\frac{\partial^2 (\nabla\cdot\mbox{\boldmath$B$})}{\partial t^2} + \frac{c_h^2}{c_p^2}\frac{\partial (\nabla\cdot\mbox{\boldmath$B$})}{\partial t} - c_h^2 \nabla^2(\nabla\cdot\mbox{\boldmath$B$})=0,
\end{eqnarray}
where $c_h$ and $c_p$ are constants, and characterize the propagation speed and the damping rate, respectively.

Equation (\ref{eq:psi}) has a semiconservative form and thus its numerical solution can benefit from the same high-order scheme applied to the original MHD equations. Let us consider the corresponding equations:

\begin{equation}
\label{eq:dedner}
 \frac{\partial}{\partial t} \left(
 \begin{array}{c}
 B_s\\
 \psi\\
 \end{array}
 \right) + 
 \left(
 \begin{array}{cc}
 0 & 1 \\
 c_h^2 & 0 \\ 
 \end{array}
 \right)
 \left(
 \begin{array}{c}
 \nabla \cdot \boldsymbol{B} \\
 {\partial \psi}/{\partial s} \\
 \end{array}
 \right)
 = \left(
 \begin{array}{c}
 0\\
 -\frac{c_h^2}{c_p^2}\psi\\
 \end{array}
 \right).
\end{equation}
The system equations without the source term on the right-hand side in each dimension are decoupled from the MHD equations, and have eigenvalues of $\lambda_{1,9} = \mp c_h$ and the right eigenvectors of $(\begin{array}{cc}
1  & \pm c_h \end{array})^T$. As $c_h$ is a constant, they are linear equations. Thus, the numerical flux at a cell surface is given as
\begin{equation}
\boldsymbol{F}_s^*=
\left(
 \begin{array}{c}
 \psi^*\\
 c_h^2B_s^*
 \end{array}
 \right) =
 \left(
 \begin{array}{c}
 \frac{1}{2}(^R \psi+^L \psi)-\frac{c_h}{2}(^R B_s-^L B_s) \\
 \frac{c_h^2}{2}(^R B_s+^L B_s)-\frac{c_h}{2}(^R \psi- ^L \psi)
 \end{array}
 \right),
\end{equation}
where $^L \psi$, $^L B_s$, $^R \psi$, and $^R B_s$ denote the left (L) and the right (R) states at the cell surface for each quantity, respectively. These can be obtained in the same manner as in the reconstruction step. The obtained numerical flux is added to the numerical flux from the HLLD Riemann solver.

The contribution from the source term $(\begin{array}{cc} 0  & -\psi c_h^2/c_p^2 \end{array})^T$ is then added in an operator-split fashion as
\begin{equation}
\psi^{(m)} = C_1^{(m)}\psi^{(0)}+C_2^{(m)}\left(\psi^{(m-1)}+\Delta t^n L_{FV}(\psi^{(m-1)})\right)\exp{\left(-\frac{c_h^2}{c_p^2}\Delta t^n\right)},
\end{equation}
where $C_1$ and $C_2$ are the weighting coefficients, $L_{FV}$ represents the finite volume differentiation in equation (\ref{eq:dedner}), and the superscripts on $\psi$, $C_1$, and $C_2$ denote the Runge--Kutta substage ($m$) in equations (\ref{eq:rk0})--(\ref{eq:rk3}).

The constants of $c_h$ and $c_p$ can be arbitrary defined. As $c_h$ represents the propagation speed of numerical errors, it is typically determined from the CFL condition as
\begin{equation}
c_h = \min_{i,j,k}\left(\Delta_i,\Delta_j,\Delta_k\right)\frac{\sigma_c}{\Delta t^n}.
\end{equation}
Then, $c_p$ is determined from the relation $c_p^2/c_h=const.$ so as to equal the transport and decay time scales. Numerical experiments have shown that $c_p = \sqrt{0.18c_h}$ results in the best performance regardless of grid resolution \citep{Dedner2002} and was adopted in CANS+.

\subsection{Source terms}
The source terms on the right-hand side of equations (\ref{eq:momentum}) -- (\ref{eq:energy}) are evaluated at the cell center at each Runge--Kutta substage. $(\eta \boldsymbol{j})$ and $(\eta \boldsymbol{j}\times\boldsymbol{B})$ in the resistivity terms are first evaluated at the cell center by the second-order finite difference for the current density $\boldsymbol{j}=\nabla\times\boldsymbol{B}$, then they are added as a numerical flux at the cell surface by the arithmetic mean of the two neighboring cell-center values. Therefore, the resistivity terms degrade the overall spatial accuracy because of its second-order representation (more details are shown in Appendix \ref{diffusion_test}).

\subsection{Some prescriptions for numerical stability}
\label{stability}
Higher-order, multidimensional codes usually suffer from vanishing and negative values of the scalar variables ($\rho$ and $p$). CANS+ ensures numerical stability by examining the following prescriptions. 
\begin{itemize}
\item In the reconstruction step, the first-order interpolation is applied for the scalar variables and the normal component of the vector fields:
\begin{itemize}
\item if one of the scalar variables recovered from the interpolated characteristic variables at a cell surface resulted in a negative value, or
\item if one of the scalar variables at a cell centered in the five-point stencil was two orders of magnitude smaller than the other cell values.
\end{itemize}
\item In the conversion step, the thermal pressure given by equation (\ref{eq:pressure}) is evaluated every time in the Runge--Kutta substage. If the pressure becomes negative, it is overwritten by the value in the previous substage. The pressure is also evaluated in terms of the plasma beta ($\beta=2p/B^2$) so as to bound the minimum $\beta$ value. CANS+ allows $\beta \geq \beta_{min} = 0.001$. The total energy density is then updated by the new pressure value. Thus, it is not strictly conserved in CANS+, while the mass conservation is satisfied.
\end{itemize}
These operations can provide stable solutions in rarefied and low-$\beta$ plasma regions as demonstrated in the two-dimensional (2D) simulations of the Parker instability as will be shown in \ref{parker}.

\subsection{CANS+ in cylindrical coordinates}
\label{cylindrical}
CANS+ in cylindrical coordinates ($R, \phi, z$) bases on the equations in a semi-conservative form as

\begin{equation}
\frac{\partial}{\partial t} 
\left(
 \begin{array}{c}
 \rho\\
 e \\
 \psi
 \end{array}
\right) + 
\left(
 \nabla_{\rm cyl} \cdot 
\left(
\begin{array}{ccc}
RF_R^{\rho} & RF_R^{e} & RF_R^{\psi}\\
F_\phi^{\rho} & F_\phi^{e} & F_\phi^{\psi}\\
F_z^{\rho} & F_z^{e} & F_z^{\psi}
\end{array}
\right)
\right)^T
= \boldsymbol{S}^{(\rho,e,\psi)}
 \label{eq:cyl1}
\end{equation}

\begin{equation}
\label{eq:cyl2}
 \frac{\partial}{\partial t} \left(
 \begin{array}{c}
 \rho v_R\\
 \rho R v_\phi\\
 \rho v_z \\
 \end{array}
 \right) 
 +\left( \nabla_{\rm cyl} \cdot
 \left(
 \begin{array}{ccc}
 R F_{RR}^{\rho v} & R^2 F_{R\phi}^{\rho v} & R F_{Rz}^{\rho v} \\
 F_{\phi R}^{\rho v} & F_{\phi\phi}^{\rho v} & F_{\phi z}^{\rho v}\\
 F_{zR}^{\rho v} & F_{z\phi}^{\rho v} & F_{zz}^{\rho v} \\
 \end{array}
 \right)
 \right)^T
 = \left(
 \begin{array}{c}
 \frac{F_{\phi\phi}^{\rho v}}{R}\\
  0 \\
  0 \\
 \end{array}
 \right)+\boldsymbol{S}^{\rho v}
\end{equation}

\begin{equation}
\label{eq:cyl3}
 \frac{\partial}{\partial t} \left(
 \begin{array}{c}
 B_R\\
 B_\phi\\
 B_z \\
 \end{array}
 \right) + 
 \left(
 \begin{array}{c}
 \frac{\partial F_{RR}^B}{\partial R} + \frac{\partial F_{\phi R}^B}{R\partial \phi}+\frac{\partial F_{zR}^B}{\partial z} \\
 \frac{\partial F_{R\phi}^B}{\partial R} + \frac{\partial F_{\phi \phi}^B}{R\partial \phi}+\frac{\partial F_{z\phi}^B}{\partial z}\\
 \frac{\partial (RF_{Rz}^B)}{R\partial R} + \frac{\partial F_{\phi z}^B}{R\partial \phi} + \frac{\partial F_{zz}^B}{\partial z}
 \end{array}
 \right)
 =\boldsymbol{S}^{B}
\end{equation}
where 
\begin{equation}
\nabla_{\rm cyl} = \left(
\begin{array}{ccc}
 \frac{\partial}{R\partial R} & \frac{\partial}{R\partial \phi} & \frac{\partial}{\partial z}
\end{array}
\right),
\end{equation}
$\boldsymbol{F}^{U}$ and $\boldsymbol{S}^U$ are the flux and the source term (equation (\ref{eq:source})) corresponding to each conservative variable ($U=\rho, \rho\boldsymbol{v}, \boldsymbol{B}, e, \psi$).

The conservation equations are discretized in a form, for example,
\begin{eqnarray}
\frac{\partial \rho_{(i,j)}}{\partial t} &=& -\frac{R_{i+1/2}F_{R(i+1/2,j,k)}^{*\rho}-R_{i-1/2}F_{R(i-1/2,jk)}^{*\rho}}{R_i\Delta R_i} \\ \nonumber
                                         &&  -\frac{F_{\phi(i,j+1/2,k)}^{*\rho}-F_{\phi(i,j-1/2,k)}^{*\rho}}{R_i\Delta \phi_j} \\ \nonumber
                                         &&  -\frac{F_{z(i,j,k+1/2)}^{*\rho}-F_{z(i,j,k-1/2)}^{*\rho}}{\Delta z_k}.
\end{eqnarray}
The source term inherent in a curvilinear coordinate system (the first term on the right-hand side of equation (\ref{eq:cyl2})) is evaluated by taking the arithmetic mean of cell-surface values as $(F_{\phi\phi(i,j-1/2)}^{\rho v}+F_{\phi\phi(i,j+1/2)}^{\rho v})/2$. This source term also degrades the overall spatial accuracy because of its second-order representation.

Special care is taken in the MP5 reconstruction step for the code in cylindrical coordinates. \cite{Mignone2014} showed that incorporating the curvature of the cell into the piecewise polynomial reconstruction, namely, the volume-weighted reconstruction, greatly improved the solutions near the origin of the coordinate axis (along the $z$-axis in cylindrical coordinates). This technique has been implemented in the cylindrical version of the code (more details are shown in Appendix \ref{appendix_cyl}).

The cylindrical version of the code has been used in global simulations of accretion disks, as shown in Section \ref{gMRI}.

\section{Code Assessments}
In this section, we present results from numerical tests to assess CANS+ by comparing with results from the second-order MUSCL scheme with the monotonized central (MC) limiter and the second-order SSP--RK scheme \citep{Gottlieb1998}. For specific heat ratio, gravity, and resistivity, $\gamma=5/3$, $\boldsymbol{g}=0$, and $\eta=0$ were adopted in the following tests unless otherwise stated.

\subsection{One-dimensional problems}
\subsubsection{Alfv\'en wave propagation}
The spatial resolution of the MP5 scheme in CANS+ was verified by 1D tests of circularly polarized Alfv\'en wave propagations with various numbers of computational cells per wavelength. We initially set the magnetic and velocity profiles as
\begin{eqnarray}
B_x &=& B_0, \nonumber\\
B_y &=& 0.1B_0\cos\left(2\pi x\right), \nonumber\\
B_z &=& 0.1B_0\sin\left(2\pi x\right), \nonumber\\
V_y &=& -B_y, \nonumber\\
V_z &=& -B_z, \nonumber
\end{eqnarray}
where we have used units of the system size $L_x=1$, the background Alfv\'en speed $V_{A0}=1$, and the Alfv\'en transit time $T_0=L_x/V_{A0}$. The CFL number $\sigma_c=0.05$ was adopted in the following analyses on spatial resolutions to minimize errors from the time integration. Otherwise, the solution becomes the third order of the SSP--RK scheme under fixed CFL conditions.

\begin{figure}
  \begin{center}
  \includegraphics[scale=0.8]{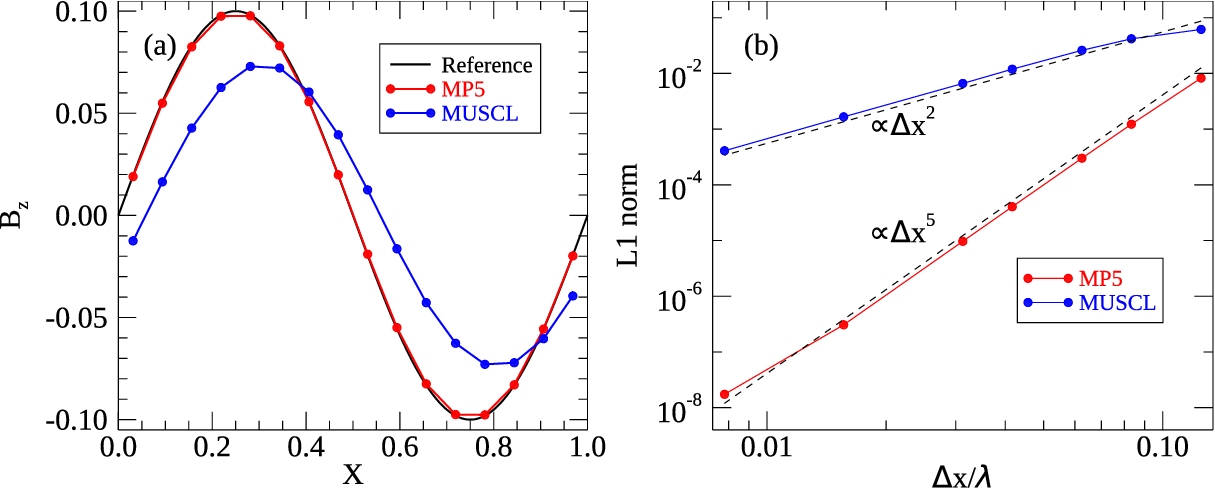}
  \end{center}
  \caption{Circularly polarized linear Alfv\'en wave propagation tests from MP5 (red) and MUSCL (blue) schemes. (a) $B_z$ profiles after five Alfv\'en transit times (t=5). The black curve indicates the analytic position of the wave. (b) $L_1$ norm errors obtained from numerical experiments with different grid resolutions. Dashed lines indicate the order of accuracy in space.}
  \label{fig:CPA}
\end{figure}

Figure \ref{fig:CPA}(a) shows spatial profiles of $B_z$ obtained by the MP5 (red) and the MUSCL scheme (blue). The wavelength is resolved with 16 cells. After five Alfv\'en transit times ($t=5$), while the amplitude of the wave has decreased by 25\% in the MUSCL scheme, the profile from the MP5 scheme almost overlapped the theoretical profile (black). This strong damping of the wave is an inherent property of the TVD schemes. In contrast, the MP5 scheme can detect smooth profiles and thus give a high-resolution result without modifying the original profile.

Numerical experiments with various cell widths gave orders of accuracy of the schemes. Figure \ref{fig:CPA}(b) shows $L_1$ norms of the numerical errors from the MP5 (red) and the MUSCL (blue) schemes. The numerical errors increased with cell width according to the degree of the interpolation polynomial of each scheme. As expected, the dashed lines indicate that the MUSCL and the MP5 schemes have the second- ($\propto \Delta x^2$) and fifth-order ($\propto \Delta x^5$) accuracy in space, respectively. 

\subsubsection{Shock tube problem}
The monotonicity preservation of CANS+ was verified by a standard shock tube problem \citep{Brio1988,Ryu1995}. We initially set for this shock tube problem the values $(\rho, p, V_x, V_y, V_z, B_x, B_y, B_z)_L=(1,1,0,0,0,0.75,1,0)$ in a region on the left-hand side of the simulation domain, $0\le x<0.5$, and $(\rho, p, V_x, V_y, V_z, B_x, B_y, B_z)_R=(0.125,0.1,0,0,0,0.75,-1,0)$ on the right-hand side in $0.5\le x\le1$ with $\gamma=5/3$ \citep{Ryu1995}. We used 512 computational cells for this problem.

\begin{figure}
  \begin{center}
  \includegraphics[scale=0.6]{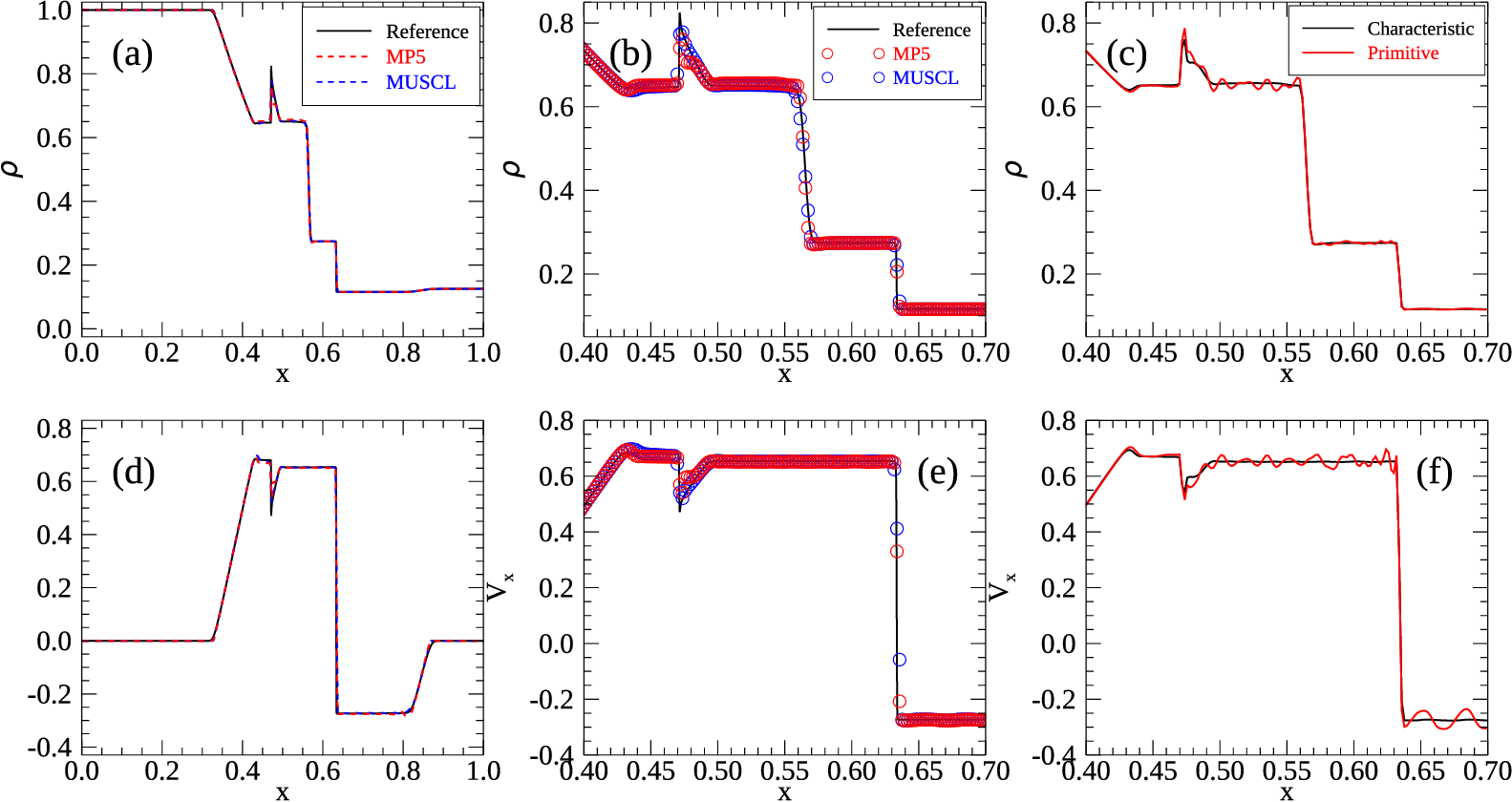}
  \end{center}
  \caption{Brio \& Wu's shock tube problem: (a) and (b) show the mass density and (d) and (e) the $x$-component of velocity profiles for the MP5 (red) and MUSCL (blue) schemes. The reference result (black) is obtained by the first-order scheme using 8192 cells. Enlarged views of the mass density and velocity profiles are shown in (b) and (e), respectively. (c) and (f) compare the differences in mass density and velocity profiles arising from the MP5 reconstructions based on the characteristic (black) and primitive (red) variables.}
  \label{fig:shktb}
\end{figure}

Figures \ref{fig:shktb}(a) and \ref{fig:shktb}(d) respectively show $\rho$ and $V_x$ profiles at $t=0.1$ obtained by CANS+ (red), and we compared them with the MUSCL scheme (blue) and the reference result obtained by the first-order scheme with 8192 cells (black). Characteristic profiles (from left to right: fast rarefaction, slow compound, contact discontinuity, slow shock, and fast rarefaction) were reproduced by CANS+. In particular, the contact discontinuity ($x\sim0.56$) and the slow shock wave ($x\sim0.63$) were resolved more sharply (figures \ref{fig:shktb}(b) and \ref{fig:shktb}(e)) than in the MUSCL scheme. A staircasing profile around the compound wave can be found when the MP5 scheme was combined with the HLLD Riemann solver. A smoother profile was obtained with the HLL solver (not shown) as previously reported with the global Lax-Friedrichs scheme \citep{Mignone2010}.

Figures \ref{fig:shktb}(c) and \ref{fig:shktb}(f) compare the results from the MP5 scheme with different variables for the interpolation to cell surfaces. The reconstruction of the characteristic variables (black) that was used in CANS+ resulted in the best performance to capture the discontinuities among other variables used for the reconstruction, including the primitive variables shown in the figure (red). Interpolation of the characteristic variables, which were actually transported by the waves, was necessary for obtaining nonoscillatory results with the present fifth-order scheme.

\subsection{Two-dimensional problems}
\subsubsection{Alfv\'en wave propagation}
\label{2DCPA}
The capability of CANS+ is demonstrated with the circularly polarized Alfv\'en wave propagation in two dimensions. By using various numbers of computational cells per wavelength, we verified the actual spatial resolution in multi dimensions. We initially set the magnetic and velocity profiles as
\begin{eqnarray}
B_l &=& B_0, \nonumber\\
B_{t1} &=& 0.1B_0\cos\left(2\pi l/\lambda\right), \nonumber\\
B_{t2} &=& 0.1B_0\sin\left(2\pi l/\lambda\right), \nonumber\\
V_{t1} &=& -B_{t1}, \nonumber\\
V_{t2} &=& -B_{t2}, \nonumber
\end{eqnarray}
where $\lambda=1$ is the wavelength, $l$ is the coordinate in the direction of the wave propagation, and $t1$ and $t2$ are the transverse components. $x$-, $y$-, and $z$-components of the magnetic field (velocity) are thus initially given as
\begin{eqnarray}
B_x &=& B_l \cos(\theta) - B_{t1} \sin(\theta), \nonumber \\
B_y &=& B_l \sin(\theta) + B_{t1} \cos(\theta), \nonumber \\
B_z &=& B_{t2}, \nonumber
\end{eqnarray}
where $\theta$ is the propagation angle with respect to the $x$-axis. The simulation box sizes are $L_x=1/\cos(\theta)$ and $L_y=1/\sin(\theta)$. The CFL number $\sigma_c=0.01$ was adopted to minimize errors from time integrations in the following analyses on spatial resolutions. We calculated numerical errors from the analytical solution after five Alfv\'en transit times along the propagation axis. In the following, results only with $\theta=30\degree$ are shown; however, we have confirmed that the resolution properties are similarly obtained with different propagation angles (see also Appendix \ref{appendix_crt}).

\begin{figure}
  \begin{center}
  \includegraphics[scale=0.7]{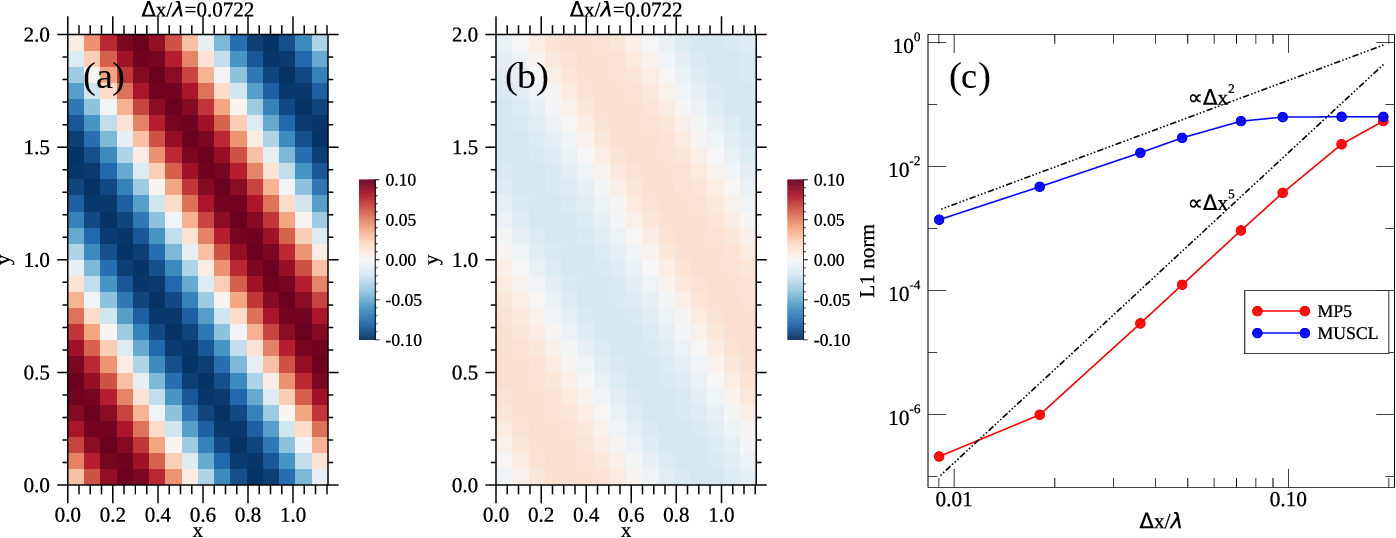}
  \end{center}
  \caption{Circularly polarized linear Alfv\'en wave propagation tests in the oblique propagation. $B_z$ profiles after five Alfv\'en transit times are shown for (a) MP5 and (b) MUSCL schemes. The red-blue colorbar is scaled to the initial amplitude. (c) $L_1$ norm errors obtained from numerical experiments with different grid resolutions for the MP5 (red) and MUSCL (blue) schemes. Dashed lines indicate the order of accuracy in space.}
  \label{fig:CPA2D}
\end{figure}

Figures \ref{fig:CPA2D}(a) and \ref{fig:CPA2D}(b) show spatial profiles of $B_z$ obtained by the MP5 and MUSCL schemes, respectively. The wavelength is resolved with 16 cells in each direction. After five Alfv\'en transit times ($t=5$), while the wave amplitude significantly reduced because the MC limiter is used in the MUSCL scheme, the initial wave amplitude was approximately retained when using the MP5 scheme with this spatial resolution. 

Numerical experiments with various cell widths in figure \ref{fig:CPA2D}(c) again shows the high capability of the MP5 scheme in two dimensions. The errors from the MP5 scheme decreased as the cell's size gets smaller following a slope expected from the fifth order interpolation. It is also worth mentioning that the numerical errors are much larger in the MUSCL scheme than in the MP5 scheme when compared at the same cell size (figure \ref{fig:CPA2D}(c)). In other words, the MP5 scheme can provide numerical solutions at the same accuracy with much coarser cells than those required in the MUSCL scheme. This result has great benefits for saving numerical costs because they increase as $\Delta x^{-(n+1)}$ in {\it n}-dimensional simulations, while the number of numerical operations in the MP5 scheme increases by a factor of two from those in the MUSCL scheme (see subsection \ref{performance}).

\subsubsection{The Kelvin--Helmhotlz instability}
Evolutions of the Kelvin--Helmholtz (K--H) instability in inhomogeneous plasma are presented as a multidimensional problem. We initially set the velocity shear profile as $V_x = -0.5V_0\tanh (y/\lambda)$, where $V_0$ was 1.6 times the Alfv\'en speed $V_A$, $\lambda$ was the half thickness of the velocity shear layer, and the mass density profile was set as $\rho = 0.5\rho_0\left[ (1-\alpha)\tanh\left(y/\lambda\right)+1+\alpha \right]$ with $\alpha = 0.1$. The magnetic field has only an out-of-plane component $\boldsymbol{B}=(0,0,B_0)$. The system size in the $x$-direction ($L_x$) is equal to the wavelength of the fastest growing mode ($L_x=11.2\lambda$) obtained by the linear analysis for the present initial conditions \citep{Matsumoto2010}. The system size in the $y$-direction ($L_y$) was in the range $-10\lambda \le y \le 10 \lambda$ ($L_y=20\lambda$).  With these initial configurations, we added a sinusoidal perturbation to the y component of the velocity inside the shear layer as $\delta V_y = 0.01V_0\sin(2\pi x/L_x)\cosh^{-2}(y/\lambda)$. Quantities were normalized to $V_A$, $B_0$, $\rho_0$, $\lambda$, and $\lambda/V_0$. The computation domain was covered by $360 \times 640$ cells with $\lambda$ being resolved by 32 cells. This initial configuration allowed us to benchmark the code's capability of capturing hydrodynamical turbulence along with the contact discontinuity.

\begin{figure}
  \begin{center}
  \includegraphics[scale=0.7]{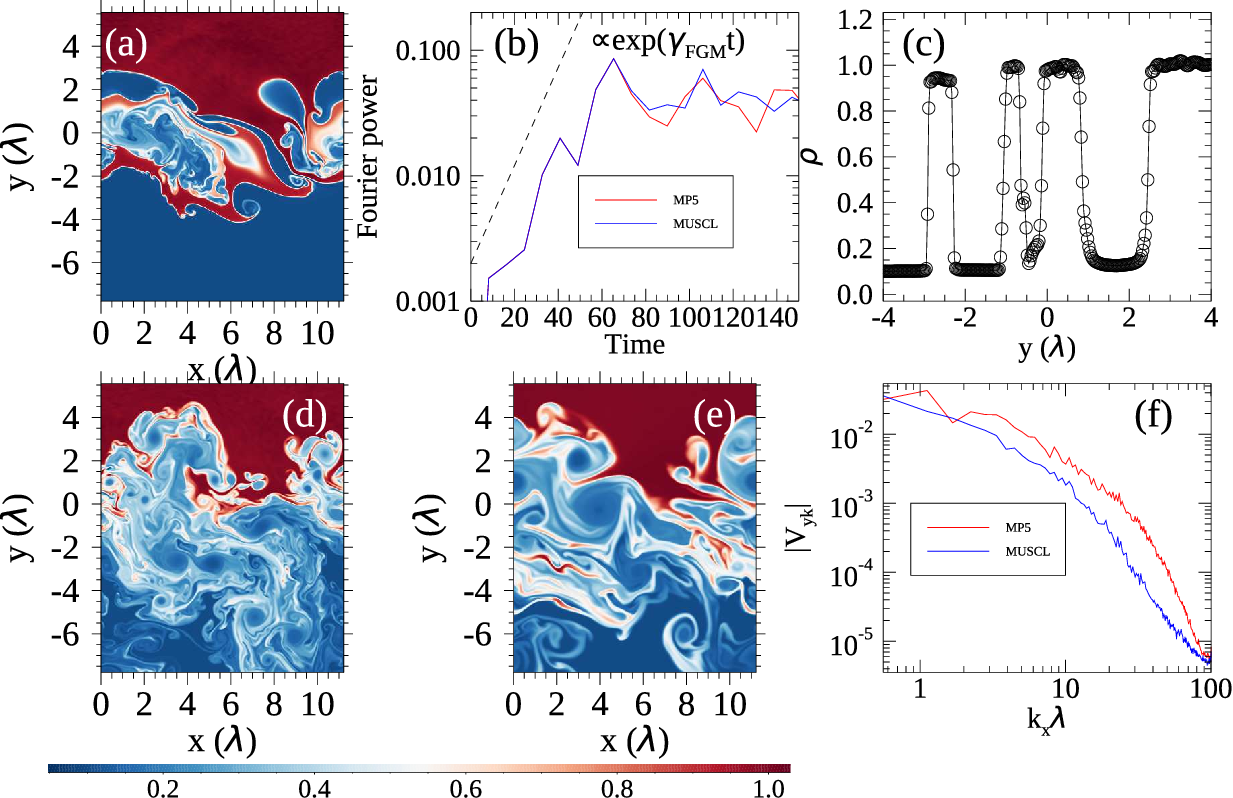}
  \end{center}
  \caption{Evolutions of the Kelvin--Helmholtz instability. (a) Mass density profile in the late linear growth phase at $t=55$. (b) Time evolutions of the Fourier amplitude of the fastest growing mode of $V_y$ from the MP5 (solid line) and MUSCL (dash--dot line) schemes. The dashed line indicates the linear growth rate obtained by \citet{Matsumoto2010}. (c) Mass density profile from the MP5 run in the $y$-direction at $x=8.4$, $t=55$. Mass density profiles from the (d) MP5 and (e) MUSCL schemes after the transition to turbulence at $t=95$. (f) Spectra of the y component of the velocity as a function of the wavenumber in the $x$-direction averaged over the simulation domain at $t=95$ for the MP5 (solid line) and MUSCL (dash--dot line) schemes.}
  \label{fig:KH}
\end{figure}

Figure \ref{fig:KH} shows the mass density profiles taken during nonlinear evolution with the MP5 (figures \ref{fig:KH}(a) and \ref{fig:KH}(d)) and the MUSCL (figure \ref{fig:KH}(e)) schemes. Saturated K--H billows naturally formed sharp contact discontinuities, which eventually broke into turbulence via the secondary K--H and Rayleigh--Taylor instabilities \citep{Matsumoto2004}. The fastest growing mode (FGM) grew exponentially at a rate of $\gamma_{FGM} = 0.09\lambda/V_0$ as expected from the linear analysis (figure \ref{fig:KH}(b)) in both results from the MP5 (solid) and MUSCL (dash--dot) schemes, whereas different behaviors in the nonlinear stage can be found in $t>70$. A sharp mass density profile at $x=8.4$ in the $y$-direction at $t=55$ (figure \ref{fig:KH}(c)) verified that the MP5 scheme was capable of capturing the contact discontinuity and development of small-scale eddies at the same time precisely. Further development of the K--H turbulence contrasted the overall mixing efficiency between different schemes. The MP5 scheme allowed the kinetic energy to be cascaded to small-scale vortices (figures \ref{fig:KH}(d) and  \ref{fig:KH}(f)), whereas the MC limiter adopted in the MUSCL scheme dissipated the cascaded energy at scales much larger than the cell size (figures \ref{fig:KH}(e) and \ref{fig:KH}(f)).

\subsubsection{The Orszag--Tang vortex}
We present results of the so-called Orszag--Tang vortex problem \citep{Orszag1979} to test the code's capability of capturing shock--shock interactions and turbulence. We initiated the problem with $(\rho, p, V_x, V_y, V_z, B_x, B_y, B_z)=(\gamma^2,\gamma,-\sin(x),\sin(x),0,-\sin(y),\sin(2x),0)$ in the simulation domain of $0\le x,\ y \le 2\pi$. The periodic boundary condition was applied in the $x$- and $y$-directions. $200 \times 200$ cells were used to compare the results from the MP5 and MUSCL schemes. Results with $800 \times 800$ cells by the MUSCL scheme are also given as a reference run.

\begin{figure}
  \begin{center}
  \includegraphics[scale=0.8]{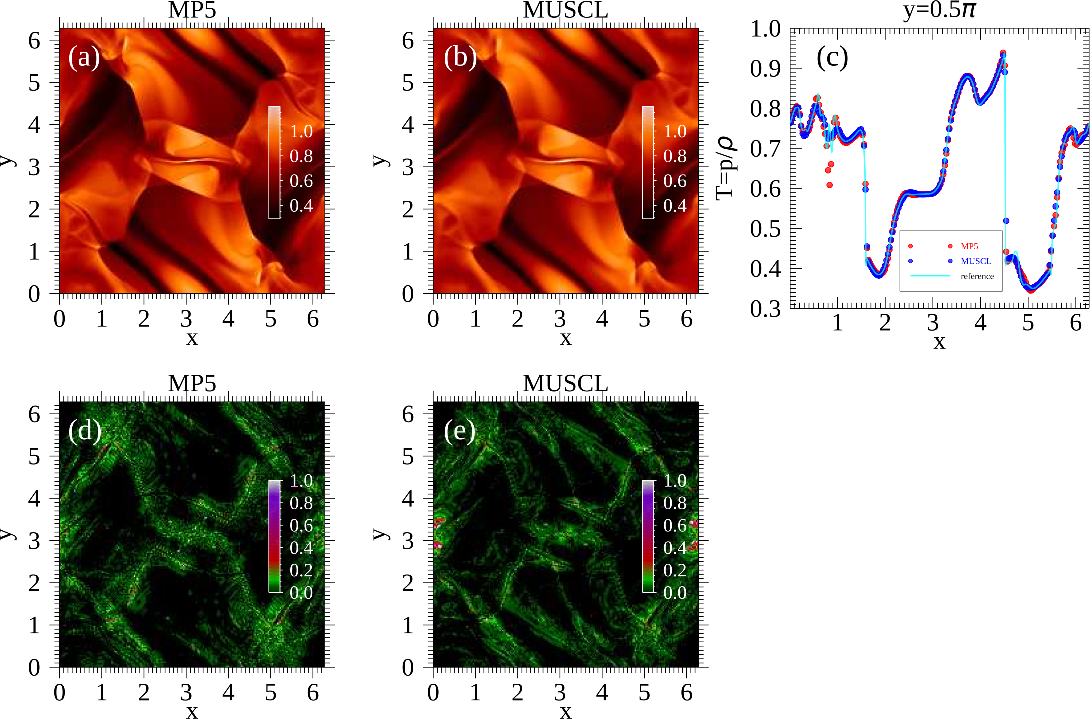}
  \end{center}
  \caption{The Orsarg--Tang vortex problem. Temperature ($T=p/\rho$) profiles at $t=\pi$ are shown for the (a) MP5 and (b) MUSCL schemes. (c) The profiles in the $x$-direction at $y=0.5\pi$ are compared between the MP5 (red filled circle) and MUSCL (blue open circle) schemes along with the result from the high-resolution run ($800 \times 800$ cells) with the MUSCL scheme (cyan curve) as a reference. The normalized divergence errors of the magnetic field (equation (\ref{eq:divb_norm})) is shown for the (d) MP5 and (e) MUSCL schemes, respectively. The color scale is saturated to 1.0 in (d) and (e).}
  \label{fig:OTvortex}
\end{figure}

Figures \ref{fig:OTvortex}(a)--\ref{fig:OTvortex}(c) show temperature ($T=P/\rho$) profiles from the MP5 (figure \ref{fig:OTvortex}(a)) and the MUSCL schemes (figure \ref{fig:OTvortex}(b)). The overall structures seem to overlap with each other as characterized by sharp discontinuities at shock waves. Besides the overall structure, a closer look at the profile along the $x$-direction at $y=0.5\pi$ (figure \ref{fig:OTvortex}(c) ) shows that the MP5 scheme (red filled circles) tended to follow a fine-scale structure of the reference run ($x\sim0.75$, cyan line) but with a oscillatory profile, whereas the the smooth profile was obtained by the MUSCL scheme (blue circles).

Figures \ref{fig:OTvortex}(d) and \ref{fig:OTvortex}(e) show profiles of divergence errors of the magnetic field. Here we introduced a normalized quantity defined by
\begin{equation}
\frac{|\boldsymbol{B}\nabla\cdot\boldsymbol{B}|}{\sqrt{(\nabla\cdot\boldsymbol{F}^{*\rho v_x})^2+(\nabla\cdot\boldsymbol{F}^{*\rho v_y})^2}},
\label{eq:divb_norm}
\end{equation}
where $\boldsymbol{F}^{*\rho v_x}$ and $\boldsymbol{F}^{*\rho v_y}$ denote the numerical flux for the $x$- and $y$-components of the momentum equations, respectively, to quantify the impact of the divergence errors to the dynamics (equation (\ref{eq:divb_origin})). Because CANS+ adopts the hyperbolic divergence cleaning method, non-negligible errors (a few to $10\%$) persist locally, in particular at shock wave fronts in both schemes. The errors are occasionally comparable to or greater than the physical force in very localized regions. This caveat that the divergence errors tended to be larger at shocks must be considered in multidimensional problems.

\subsubsection{Parker instability}
\label{parker}
Magnetized plasma stratified under the gravity becomes destabilized as a result of the buoyancy force against the magnetic tension force. This instability is known as the Parker instability \citep{Parker1966}, and its nonlinear evolution is characterized by bent magnetic field lines in rarefied plasma \citep{Matsumoto1988}. We chose this problem as a benchmark test for evaluating the code's capability of solving low-$\beta$ plasma in multidimensions. In the present test, the gravity acceleration term $\boldsymbol{g}$ on the right-hand side of equation (\ref{eq:momentum}) was retained. 

The Parker instability was initialized with the gravity acceleration profile
\begin{equation}
\boldsymbol{g} = \left(0, -g_0 \tanh{\left(\frac{y}{H_g}\right)}\right),
\end{equation}
and the temperature
\begin{equation}
T = T_0+\frac{1}{2}(T_1-T_0)\left[\tanh{\left(\frac{|y|-y_0}{H_t}\right)}+1\right].
\end{equation}
The mass density was determined from the static equilibrium
\begin{equation}
\frac{d}{d y} \left[\left(1+\beta_0^{-1}\right)p\right] = \rho g,
\end{equation}
where $\beta_0$ is the ratio of the thermal pressure to the magnetic pressure, and $p = \rho T/\gamma$. The magnetic field has only the $x$-component ($B_x$) with strength given by the plasma beta $\beta_0$. The spacial length, velocity, and time were normalized to the scale height $H_0$, the sound speed $C_{s0}$ at $y=0$, and the transit time $H_0/C_{s0}$, respectively. Other quantities such as $g$, $\rho$, $p$, $T$, and $B_x$ were normalized to $C_{s0}^2/H_0$, $\rho_0$, $\rho_0C_{s0}^2$, $C_{s0}^2/(\gamma/m)$, and $\sqrt{\rho_0C_{s0}^2}$, respectively. We adopted $\gamma=1.05$, $g_0=1.47$, $H_g=5.0$, $H_t=0.5$, $y_0=10.0$, $T_0=1$, $T_1=25$, and $\beta_0=1.0$.

\begin{figure}
  \begin{center}
  \includegraphics[scale=0.6]{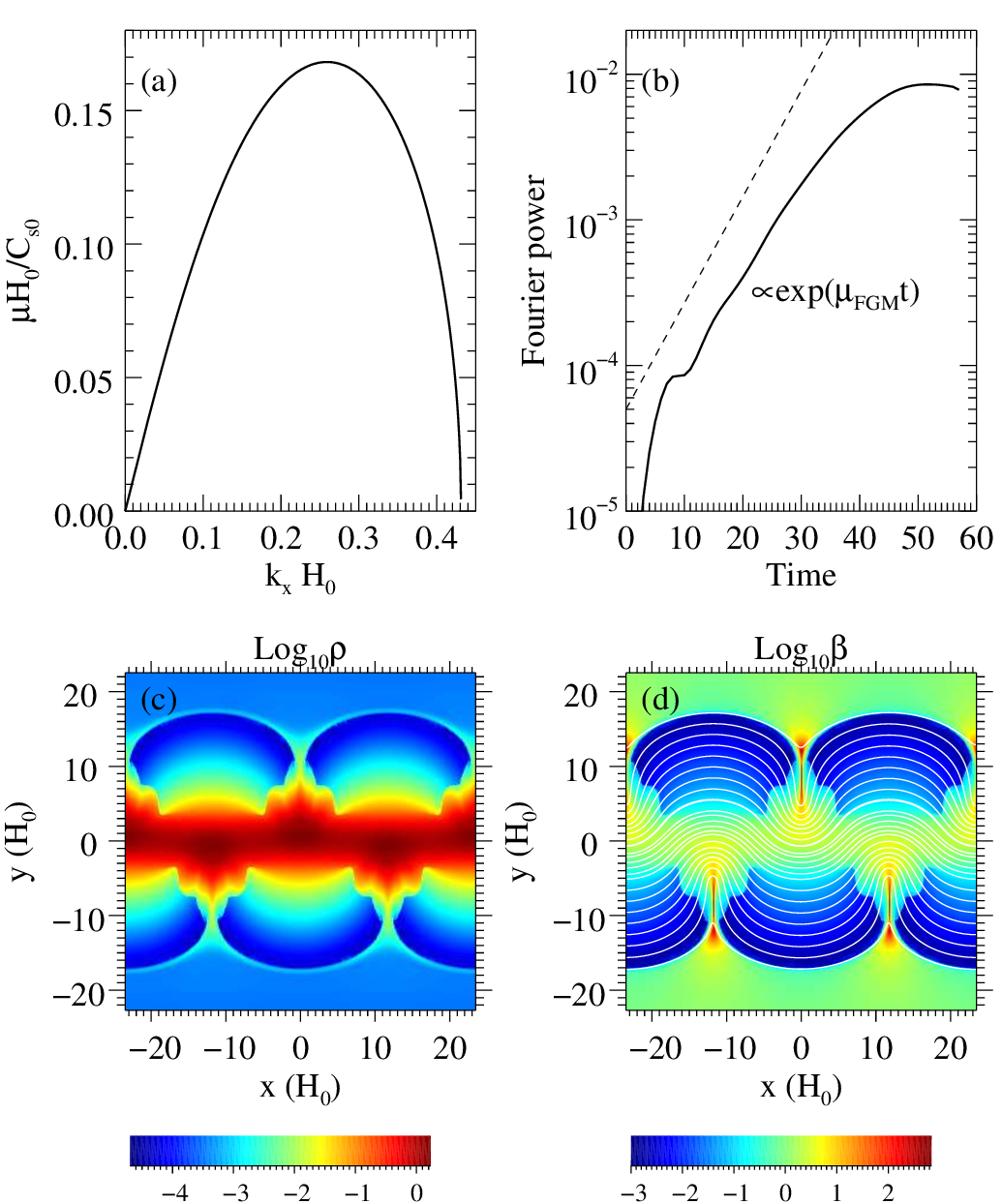}
  \end{center}
  \caption{(a) Linear growth rate of the Parker instability as a function of the wavenumber in the $x$-direction. (b) Time evolution of the Fourier power of the m=2 mode (the fastest growing mode) of $B_y$ by CANS+. The dashed line indicates the linear growth rate from the linear analysis in (a). (c) Mass density and (d) plasma $\beta$ profiles at $T=50$ in logarithmic scales. Magnetic field lines are represented by white solid lines in (d).}
  \label{fig:Parker}
\end{figure}

With these initial configurations, we added a perturbation to $V_x$ in an antisymmetric form with respect to the $y=0$ plane as
\begin{eqnarray}
 && \delta V_x= 0.05\sin\left(\frac{2\pi x}{\lambda}\right) \times \nonumber \\
 && \left[\left\{\tanh{\left(\frac{y+y_2}{H_{pt}}\right)}-\tanh{\left(\frac{y+y_3}{H_{pt}}\right)}\right\}-\left\{ \tanh{\left(\frac{y-y_2}{H_{pt}}\right)}-\tanh{\left(\frac{y-y_3}{H_{pt}}\right)}\right\}\right],
\end{eqnarray}
where $y_2=4.0$ and $y_3=1.0$ restricted the perturbation in $1.0 \le |y| \le 4.0$, and $H_{pt}=0.5$. $\lambda=7.5\pi$ was chosen so that the wavelength of the applied perturbation corresponded to the fastest growing mode of the Parker instability under the present initial conditions (figure \ref{fig:Parker}(a)). A benchmark test was conducted in a simulation domain of $-7.5\pi \le x \le 7.5\pi$ and $-15\pi \le y \le 15\pi$ with the periodic boundary condition in the $x$-direction and the free boundary condition in the $y$-direction. The scale height $H_0$ was resolved by $20/\pi$ cells corresponding to the total number of cells of $300 \times 600$.

Figure \ref{fig:Parker}(b) shows the time evolution of the Fourier amplitude of the initial perturbation of the fastest growing mode. We found that the most unstable mode grew exponentially until $t=50$ at a rate expected from the linear analysis (figure \ref{fig:Parker}(a)). 

In the nonlinear stage, magnetic loops were lifted up by the buoyancy force against the magnetic tension force. The plasma inside the loops fell down along the field lines creating high-density regions in the footpoints (figure \ref{fig:Parker}(c)). The loop-top region therefore became rarefied in the nonlinear stage. Usual finite volume methods based on the conservative form of the MHD equations eventually result in disastrous numerical solutions in such a rarefied, low-$\beta$ plasma, but CANS+ successfully solved its evolution in which the mass density and the plasma $\beta$ decreased down to $\sim 10^{-5}$ and $\sim 10^{-3}$ (figures \ref{fig:Parker}(c) and \ref{fig:Parker}(d)), respectively, in the late nonlinear stage by virtue of the prescriptions presented in subsection \ref{stability}.

\subsubsection{Magnetic reconnection}
\label{reconnection}
When magnetic field lines are imposed to form an antiparallel geometry, the magnetic field topology changes through reconnection of the field lines. Because it accompanies the magnetic energy conversion to the plasma kinetic energy, the magnetic reconnection has been studied extensively to understand explosive phenomena, such as flares in the solar corona and pulsar winds, and terrestrial substorms. It is also an important process in dynamo processes in accretion disks. The topology change during reconnection is characterized by bent magnetic field lines and bipolar trans-Alfv\'enic jets from the reconnecting ``x" point, and the resulting plasmoid evolution. It was recently found by performing high-resolution MHD simulations that various MHD shock waves and discontinuities are formed as a result of interactions between the jet and plasmas surrounding the plasmoid \citep{zenitani2011,zenitani2015}. Here we present the code's capability of capturing such structures and turbulence associated with the plasmoid evolution. To initiate the reconnection, the resistivity $\boldsymbol{\eta}$ terms on the right-hand side of equations (\ref{eq:induction}) and (\ref{eq:energy}) were retained.

We examined simulation runs following \citet{zenitani2011}, in which initial configurations were given as the Harris equilibrium: $B_x = B_0\tanh(y/\lambda)$, $B_y = B_z = 0$, $\boldsymbol{V} = 0$, $\rho = \rho_0(1+\cosh^{-2}(y/\lambda)/\beta_0)$, $p = 0.5B_0^2(\beta_0+\cosh^{-2}(y/\lambda))$, where $\lambda$ is the current sheet half-thickness. The resistivity was locally added in the simulation domain around the ``x" point as $\eta = \eta_0+(\eta_1-\eta_0)\cosh^{-2}(\sqrt{x^2+y^2}/\lambda)$. Perturbations were initially added to $B_x$ and $B_y$ components through the vector potential of $\delta A_z = 0.06B_0\lambda\exp{[-(x^2+y^2)/4\lambda^2]}$. To save computation time, we only solved one quadrant of the reconnection region by applying symmetric conditions at $x=0$ and $y=0$. The simulation domain was therefore $0 \le x \le 200\lambda$ and $0 \le y \le 50\lambda$ with $\lambda$ being resolved by 30 computational cells ($6000 \times 1500$ cells). The spacial length, velocity, and time were normalized to $\lambda$, $V_{A0}=B_0/\sqrt{\rho_0}$, and $\lambda/V_{A0}$ with $B_0=\rho_0=1.0$, $\beta_0 = 0.2$ and $\eta_1=1/60$. We set $\eta_0=0.0$ to highlight the code's capability.  

\begin{figure}
  \begin{center}
  \includegraphics[scale=0.65]{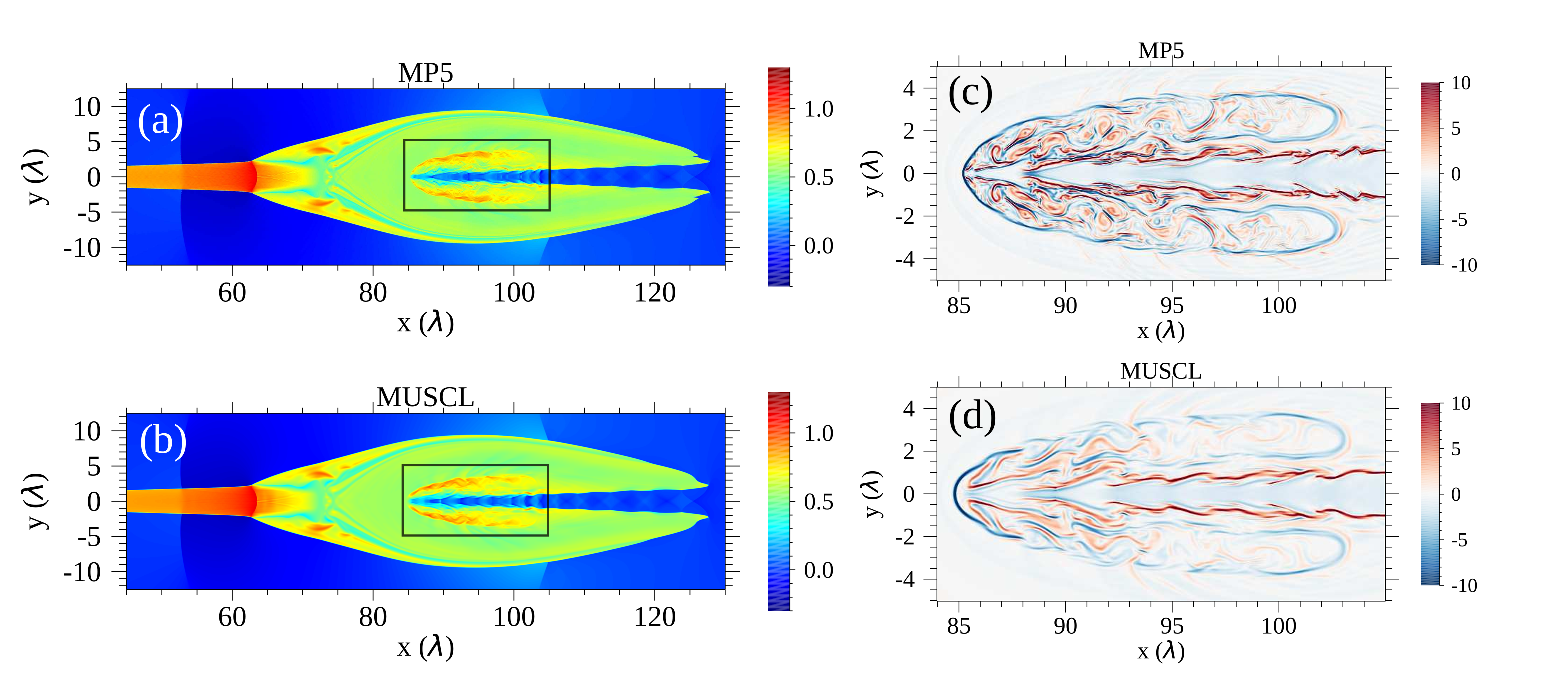}
  \end{center}
  \caption{(a) and (b) Profiles of the $x$-component of the velocity at $T=250$ for the MP5 and MUSCL schemes, respectively, characterizing a heart-shaped structure of the plasmoid downstream of the reconnection region. A mirrored image in $y<0$ is also shown for visual purposes. (c) and (d) Enlarged views of the $z$-component of the current density ($\nabla \times \boldsymbol{B}$) in the open box area in (a) and (b) for the MP5 and MUSCL schemes, respectively. The color scale is saturated at $\pm10$ to visually identify fine-scale structures inside and around the jets.}
  \label{fig:mrx}
\end{figure}

Figure \ref{fig:mrx} shows $V_x$ profiles at $t=250$ characterizing a heart-shaped plasmoid structure downstream of the reconnection region. The plasmoid is formed as a result of interaction between the collimated jet from the reconnection region ($V_x > \sim1$ in $x < 62$ and $-3 < y < 3$) and the stationary plasma in the current sheet. Vertical slow shock wave fronts in the outer ($x \sim 104$) and postplasmoid ($x \sim 53$) regions, and multiple reflections of shock waves (shock diamonds) in the current sheet ($105 \leq x \leq 127$) were clearly captured by both the MP5 and MUSCL schemes (figures \ref{fig:mrx}(a) and \ref{fig:mrx}(b)) in addition to the various MHD discontinuities and shock waves, as reported previously \citep{zenitani2011}. 

The plasmoid motion in the $x$-direction pushes the stationary plasma away from the current sheet in $85<x<88$ to $y\sim \pm2$, making the velocity shear between the swept and surrounding plasmas. This velocity profile can be a source of turbulence through excitation of the K--H instability \citep{zenitani2011}. Figures \ref{fig:mrx}(c) and \ref{fig:mrx}(d) show enlarged views of the $z$-component of the current density ($\nabla \times \boldsymbol{B}$) in the area surrounded by dashed lines in figures \ref{fig:mrx}(a) and \ref{fig:mrx}(b). In Figure \ref{fig:mrx}(c), one can see a series of oblique red lines at $90<x<100$, $y = \pm 4$.  They correspond to small shocks in front of the humps of the K--H waves.  Although the large-scale pictures looked similar in the MP5 and MUSCL runs (figures \ref{fig:mrx}(c) and \ref{fig:mrx}(d)), the MP5 scheme resolved both the shocklets and the highly turbulent current vortices much better.

\subsection{Parallel scaling and computation efficiency}
\label{performance}
Hybrid parallelization is implemented into CANS+ using the Message Passing Interface (MPI) library and OpenMP. The code has been optimized on massively parallel supercomputer systems to run effectively. We examined a parallel efficiency of the 3D code on K computer at the RIKEN Center for Computational Science, Reedbush at the Information Technology Center, the University of Tokyo, and ATERUI II at the Center for Computational Astrophysics, National Astronomical Observatory Japan.

The K computer is composed of 82,944 computation nodes, and each node has eight processor cores. The node's peak performance is 128 GFLOPS ($128\times10^9$ floating-point operations per second). For this system, intra- and internode communications were respectively realized by OpenMP directives and MPI libraries.

The Reedbush and ATERUI II systems are based on many Intel Xeon CPUs. The Reedbush system is composed of 420 computation nodes, and each node has 36 processor cores ($18$ cores $\times 2$ CPUs). The ATERUI II system is composed of 1,005 nodes each of which has 40 cores ($20$ cores $\times 2$ CPUs). The node's peak performances are 1.2 TFLOPS ($1.2\times10^{12}$ FLOPS) and 3.1 TFLOPS, respectively. For these systems, four MPI processes and eight OpenMP threads were used in each computation node.

\begin{figure}
  \begin{center}
  \includegraphics[scale=0.6]{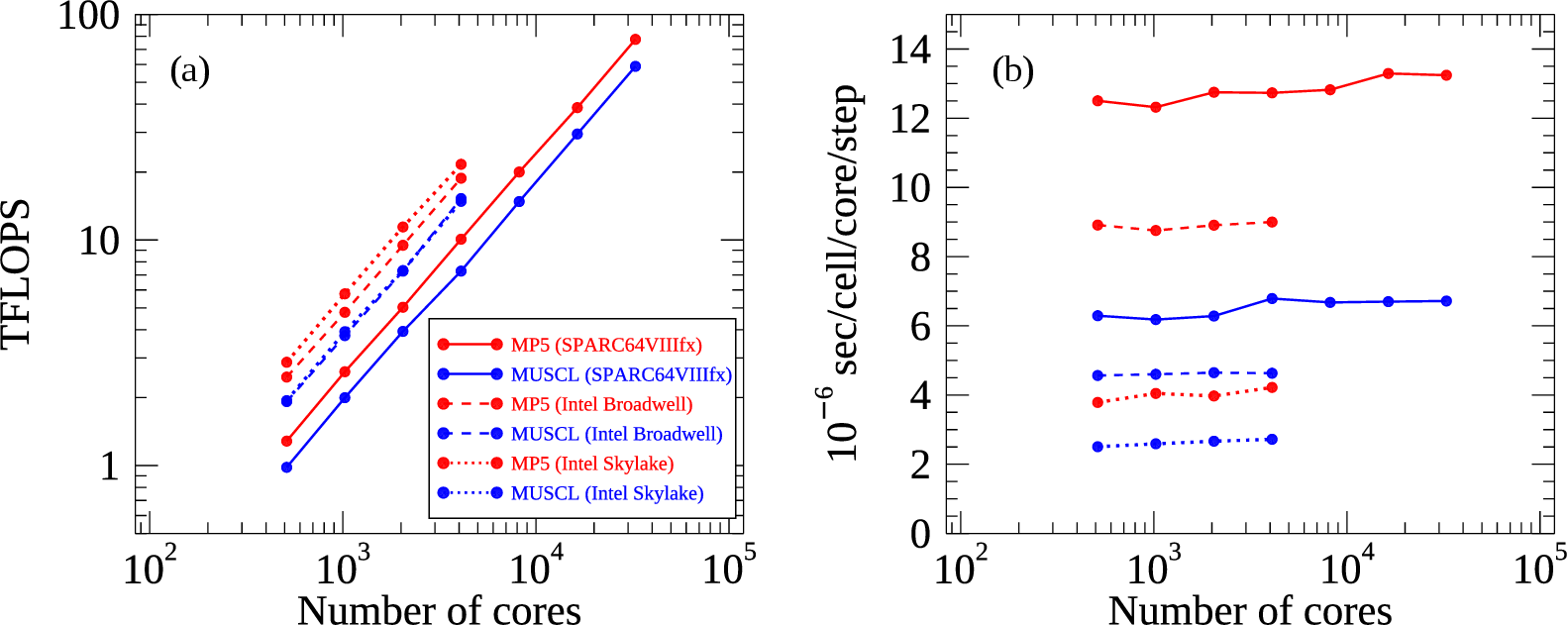}
  \end{center}
  \caption{(a) Parallel scaling of CANS+ code with different numbers of processor cores. (b) Computation speed of the code in terms of time taken for updating variables at a cell per core per time step.}
  \label{fig:performance}
\end{figure}

\begin{table}
  \caption{Computation efficiency and system's specifications}
  \begin{tabular}{|c|c|c|c|} 
\hline
     & K computer & Reedbush & ATERUI II \\
\hline
CPU & Fujitsu SPARC64VIIIfx & Intel Xeon Broadwell-EP & Intel Xeon Skylake \\
\hline
system's BF ratio & 0.50 & 0.13 & 0.08 \\
\hline
SIMD vector length (bit) & 128 & 256 (AVX2) & 512 (AVX512)\\
\hline
\begin{tabular}{c}compiler \& compiler option\end{tabular}  & \begin{tabular}{c}mpifrtpx \\-Kfast,openmp -x100\end{tabular} & \begin{tabular}{c}mpiifort \\ -O3 -xCORE-AVX2 \\ -qopenmp\end{tabular} & \begin{tabular}{c}ftn\footnotemark \\ -O3 -h omp\end{tabular} \\
\hline
\begin{tabular}{c}computation efficiency \\ to the peak performance (MP5)\end{tabular} & $15.3\%$ & $13.9\%$ & $7.2\%$ \\
\hline
\begin{tabular}{c}computation efficiency \\ to the peak performance (MUSCL)\end{tabular} & $11.6\%$ & $10.9\%$ & $4.8\%$ \\
\hline
\begin{tabular}{c}$\mu$ sec/step/core/cell (MP5)\end{tabular} & $12.8$ & $8.8$ & $4.0$ \\
\hline
\begin{tabular}{c}$\mu$ sec/step/core/cell (MUSCL)\end{tabular} & $6.5$ & $4.6$ & $2.6$ \\
\hline
\end{tabular}
\label{tab:performance}
\end{table}
\footnotetext{The vendor-provided compiler (ftn) was used because it resulted substantially better performances than the Intel compiler on this system.}

Figure \ref{fig:performance} shows a benchmark result without data I/O functions after five-minute runs (approximately $3,000$ time steps with the MP5 scheme on the K computer). The scaling was obtained with a fixed number of cells of $64 \times 16 \times 64$ per MPI process (weak scaling) by using 512--32,768 processor cores (64--4,096 MPI processes) on the K computer, 512--4,096 cores (64--512 MPI processes) on the Reedbush and the ATERUI II systems. The computation speed increased almost linearly with increasing numbers of cores up to 32,768 on all systems, achieving a parallel efficiency of $94.4\%$ for the MP5 scheme on the K computer (red solid line in Figure \ref{fig:performance}(a)). The code with the MUSCL scheme also speeds up linearly with increasing numbers of cores ($93.6\%$ efficiency, blue solid line).

With the help of the core's SIMD (single instruction multiple data) capability, CANS+ runs efficiently also with respect to the system's performance. The code with the MP5 scheme runs at $15.3\%$ on average to the peak performance of the K computer with the tested numbers of cores, whereas the code with the MUSCL scheme runs at $11.6\%$; the MP5 reconstruction is an efficient scheme in terms of floating-point operations. The performance on other systems with smaller BF ratios (ratio of the memory bandwidth to FLOPS) resulted lower efficiencies of $13.9\%$ (MP5) and $10.9\%$ (MUSCL) on the Reedbush system, and $7.2\%$ (MP5) and $4.8\%$ (MUSCL) on the ATERUI II system.

Figure \ref{fig:performance}(b) shows actual computation time during MHD simulations with various numbers of processor cores. Despite additional numerical costs including the three-stage SSP--RK time integration, the code with the MP5 scheme runs at $12.8\ \mu$s/core on average for updating the MHD variables at a cell in one time step on the K computer. This time is only twice as large as that with the MUSCL scheme with the two-stage SSP--RK integration ($6.5\ \mu$s/cell/core/step), because of the high performance in the floating-point operations with the MP5 scheme. The code runs certainly faster on newer systems of the Reedbush ($8.8$ and $4.6\ \mu$s/cell/core/step, respectively) and the ATERUI II ($4.0$ and $2.6\ \mu$s/cell/core/step, respectively). Note, however, that the ratio of the computation time of the MP5 run to the MUSCL run decreases from $12.8/6.5 \sim 2.0$ on the K computer to $4.0/2.6 \sim 1.5$ on the latest low-BF-ratio system. The performance results and corresponding system's specifications are summarized in Table \ref{tab:performance}.
 
Because the MP5 reconstruction implemented in CANS+ requires more floating-point operations per cell than those for the second-order MUSCL scheme, it is essentially a less memory-intensive code. In other words, the code takes less time to load the required data from the physical memory than that for floating-point operations. This is a good property for recent low BF ratio systems. In general, the high-resolution (greater than the fifth order) code is suitable for massively parallel, petascale to exascale supercomputer systems because of its high efficiency.

\section{Application to Global Simulations of a Black Hole Accretion Disk}
\label{gMRI}
In this section, we present global simulations of an accretion disk around a black hole as an application of CANS+. The long-term evolution was characterized by a sharp contact discontinuity between the hot, dilute corona and the cold, dense, rotating disk, compressible magnetic turbulence via the magnetorotational instability (MRI), the resulting mass accretion (advection), and the periodic dynamo process through the Parker instability \citep{Machida2013}. All of these mechanisms were successfully solved by adopting the HLLD approximate Riemann solver, the MP5 reconstruction, and the hyperbolic divergence cleaning method in CANS+.

\subsection{Initial torus model}
We examined the evolution of an accretion disk in cylindrical coordinates ($R$, $\phi$, $z$) initially given as a torus threaded by the toroidal magnetic field embedded in non-rotating, hot and dilute plasma (corona) \citep{Okada1989,Machida2003}. General relativistic effects around the black hole are modeled by the pseudo-Newtonian gravitational potential \citep{Paczynsky1980},
\begin{equation}
\Phi = -\frac{GM}{r-r_S},
\label{eq:pot}
\end{equation}
where $r$ is radial distance from the black hole in spherical coordinates, $G$ is the gravitational constant, $M$ is the black hole mass, and $r_S$ is the Schwarzschild radius. Then the gravitational acceleration was obtained by $\boldsymbol{g}=-\boldsymbol{\nabla} \Phi$. 

The torus was initially given by setting a density profile \citep{Nishikori2006} as
\begin{equation}
\rho_t = \rho_0 \left\{ \frac{\max\left[\Psi_0-\Phi-L^2/2R^2,0\right]}{K\gamma/(\gamma-1)\left(1+\beta_0^{-1}R^{2(\gamma-1)}/R_0^{2(\gamma-1)}\right)}\right\}^{1/(\gamma-1)},
\end{equation}
where $R_0$, $\rho_0$ and $\beta_0$ are the values at the center of the torus at which the mass density has a maximum value, $K=p_t/\rho_t^\gamma$ characterizes the polytropic relation between the gass pressure $p_t$ and $\rho_t$ inside the torus. $L$ is the specific angular momentum in a functional form of 
\begin{equation}
L = L_0 \left(\frac{R}{R_0}\right)^a,
\end{equation}
where $L_0$ is the value of the Keplerian flow at $R_0$ and $a$ is a constant. $\Psi_0$ is a potential energy at $R=R_0$ given by
\begin{equation}
\Psi_0 = \Phi_0+\frac{L}{2R_0^2}+\frac{\gamma K}{\gamma-1}\left(1+\frac{1}{\beta_0}\right),
\end{equation}
where $\Phi_0$ is the gravitational potential at $r=R_0$. This potential energy is a constant provided the initial toroidal magnetic field is in the form of \citep{Okada1989}
\begin{equation}
B_{\phi} = \sqrt{\frac{2K}{\beta_0}\rho_t\left(\rho_t \frac{R^2}{R_0^2}\right)^{\gamma-1}}.
\end{equation}
 The mass density for the corona was given by \citep{Nishikori2006}
\begin{equation}
\rho_c = \rho_{c0}\exp{\left(-\frac{\gamma}{C_{sc}^2}\left(\Phi-\Phi_0\right)\right)},
\end{equation}
where $C_{sc}=\sqrt{\gamma p_c/\rho_c}$ is the characteristic sound speed in the corona. The total mass density is then given by $\rho = \rho_t + \rho_c$.

\subsection{Numerical setup}
The time evolution of the black hole accretion disk was obtained by the cylindrical version of CANS+ code (see subsection \ref{cylindrical}). The spacial length, velocity, and time were normalized to the initial torus position ($R_0$), the rotating speed of the torus ($V_0=L/R_0$), and the rotational period ($t_0=2\pi R_0/V_0$), respectively, and we will discuss in units of $R_0=V_0=\rho_0=1$. In the following numerical experiments, we adopted $\gamma=5/3$, $\beta_0=100$, $r_S=0.1$, $K=0.05$, $\rho_{c0}=3\times 10^{-4}$, and $C_{sc}^2 = 5.0$. $L$ was set constant ($a=0$) inside the torus. The number of computational cells in each direction was $(N_R,\ N_\phi,\ N_z)=(512,\ 128,\ 512)$. The cell sizes in the $R$- and $z$-directions were $\Delta R=7.5\times 10^{-3}$ in $0 \le R \le 3.0$ and $\Delta z=7.5\times 10^{-3}$ in $-1.5 \le z \le 1.5$. Outside these regions, the cell sizes were increased by $5\%$ with respect to the neighboring cell size as $\Delta R_{i+1}/\Delta R_i = \Delta z_{k+1}/\Delta z_k (z>0) = \Delta z_{k-1}/\Delta z_k (z<0)= 1.05$, where $i$ and $k$ are the cell number, and are bounded by a maximum value of 0.1. The cell size in the $\phi$-direction was $\Delta \phi = 2\pi/N_\phi$. To save computation time, we set the innermost cell size to be $\Delta R_1=4\Delta R$. The computational domain consequently covered $0 \le R \le 9.4$, $0 \le \phi \le 2\pi$, and $-8.0 \le z \le 8.0$. 

The periodic boundary condition was applied in the $\phi$-direction, whereas the free boundary condition was applied in the $z$-direction and the outermost region in the $R$-direction. All physical quantities were absorbed inside the spherical region of $r \le 0.2$. This was accomplished by damping deviation of a physical quantity, $q$, from the initial state $q_0$ with a damping rate $a_i$ as \citep{Machida2003}
\begin{eqnarray}	
a_i = 0.1 \left[1.0-\tanh{\left(\frac{r-0.2+5\Delta R}{2\Delta R}\right)}\right], \\
q^{new} = q-a_i(q-q_0).
\end{eqnarray}

\subsection{Results}
\subsubsection{MP5 vs. MUSCL schemes}
\begin{figure}
  \begin{center}
  \includegraphics[scale=0.8]{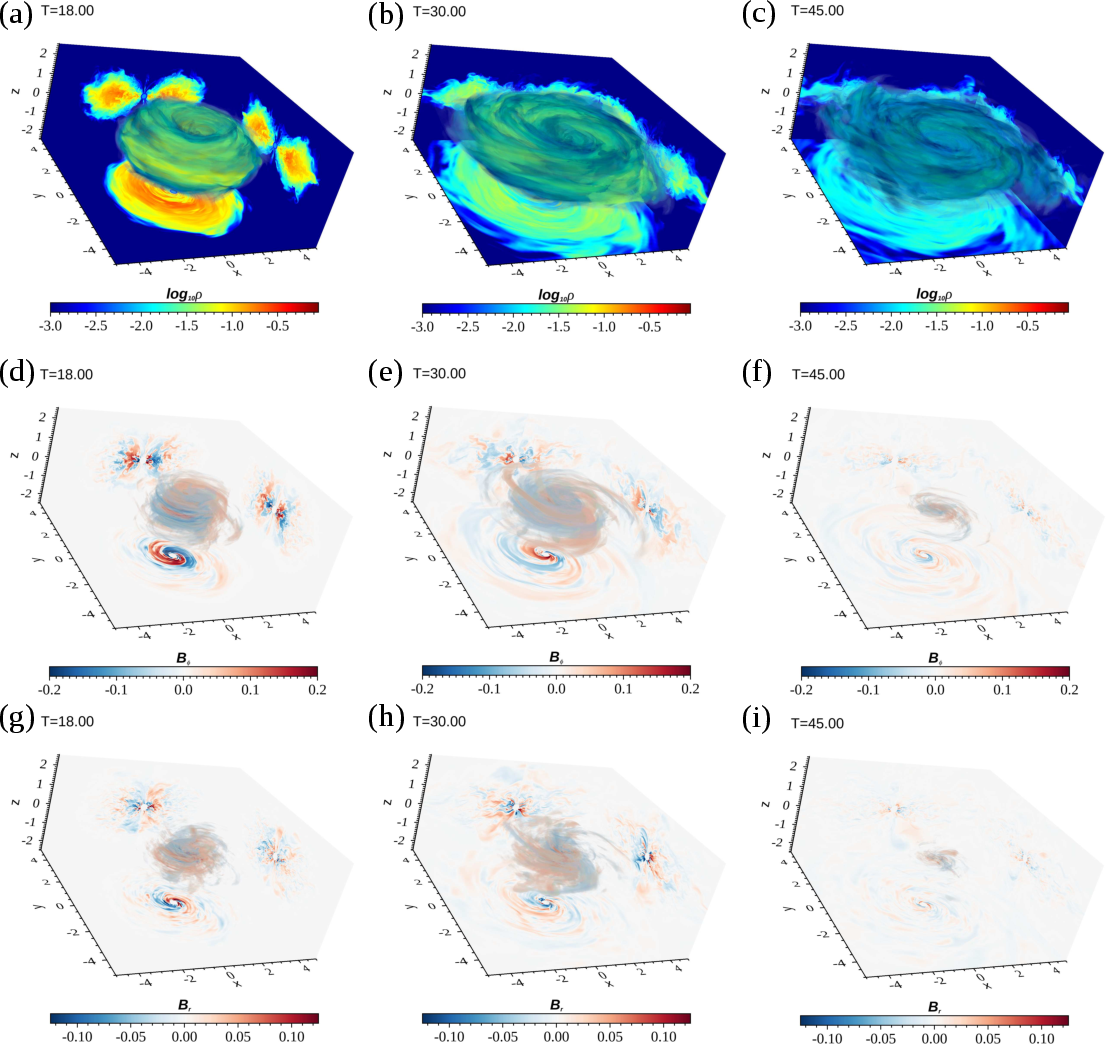}
  \end{center}
  \caption{Time evolution of the accretion disk obtained by the MP5 scheme. From left to right, snapshots at $t=18.0$ (first column), $t=30.0$ (second column), and $t=45.0$ (last column) are shown for the mass density (top row) and $\phi$- and $R$-components of the magnetic field (middle and bottom rows). Profiles at the equatorial and meridian planes passing through the black hole are projected on the $x$--$y$, $x$--$z$, and $y$--$z$ planes.}
  \label{fig:gMRI_MP5}
\end{figure}

Using the same initial setup, we compared results from the MP5 and MUSCL schemes. Figure \ref{fig:gMRI_MP5} shows the time evolution of the accretion disk solved by the MP5 scheme. Inside the torus threaded by the azimuthal magnetic field, the MRI exponentially grew (figures \ref{fig:gMRI_MP5}(a), \ref{fig:gMRI_MP5}(d), and \ref{fig:gMRI_MP5}(g)) until $t=20$. The Maxwell stress ($B_RB_\phi$) generated by the MRI (figures \ref{fig:gMRI_MP5}(d), \ref{fig:gMRI_MP5}(e), \ref{fig:gMRI_MP5}(g), and \ref{fig:gMRI_MP5}(h)) enhanced outward momentum transport, resulting in continuous mass accretion into the black hole (figures \ref{fig:gMRI_MP5}(b) and  \ref{fig:gMRI_MP5}(c)). 

\begin{figure}
  \begin{center}
  \includegraphics[scale=0.5]{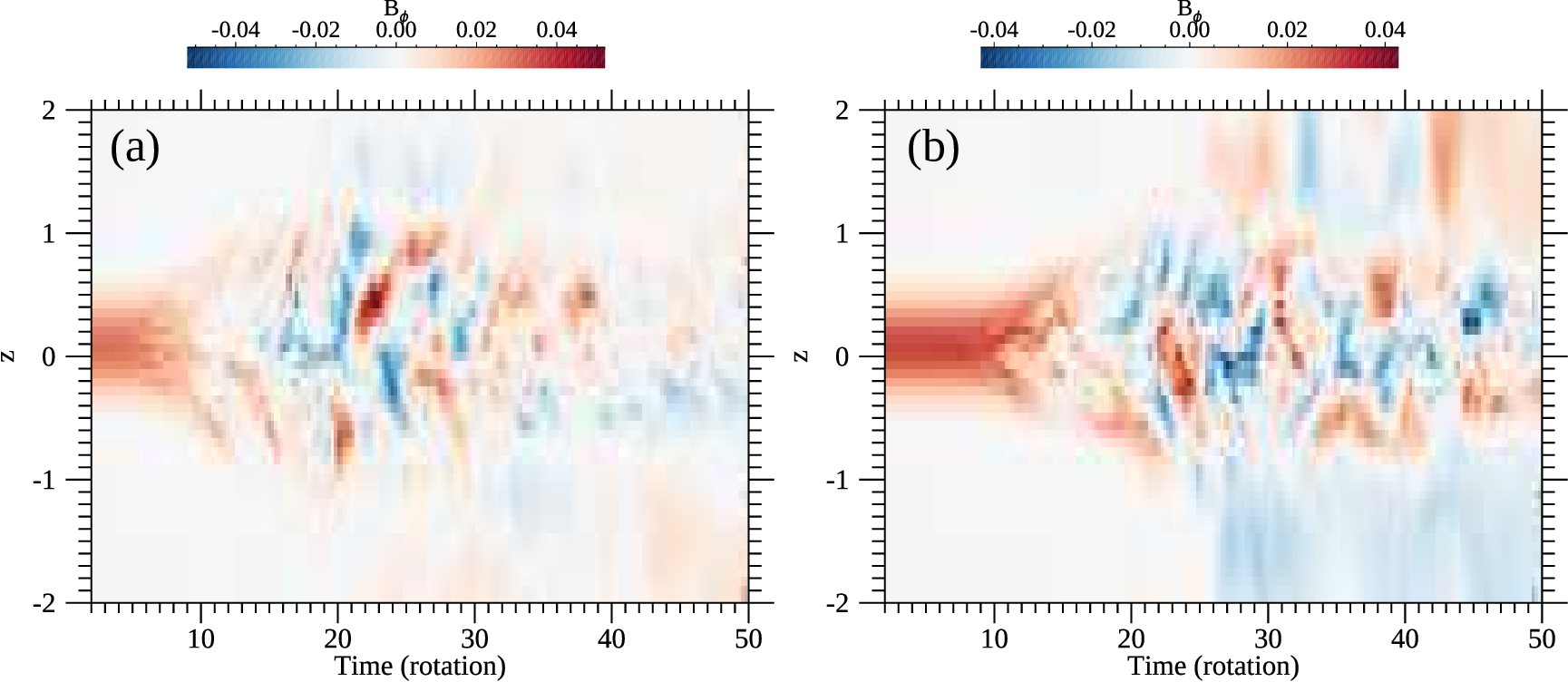}
  \end{center}
  \caption{Butterfly diagram for the azimuthal component of the magnetic field from results by the (a) MP5 and (b) MUSCL schemes. }
  \label{fig:gMRI_Butterfly}
\end{figure}

After nonlinear saturation of the MRI, the poloidal component of the magnetic field was created via the Parker instability, allowing escape of magnetic flux to the corona, which in turn caused a reversal of the sign of the azimuthal component inside the disk \citep{Machida2013}. This reversal of the toroidal magnetic field occurred periodically during its long-term evolution in the simulation run with the MP5 scheme. Figure \ref{fig:gMRI_Butterfly}(a) shows a butterfly diagram for the azimuthal component of the magnetic field averaged in the region $0.2\le r \le 0.5$ and $0\le\phi\le 2\pi$. Reversals of the sign of the magnetic field at the disk center occurred after the preceding buoyant motions of the magnetic flux due to the Parker instability. This dynamo process persisted during the simulation run up to $t\sim40$. 

\begin{figure}
  \begin{center}
  \includegraphics[scale=0.5]{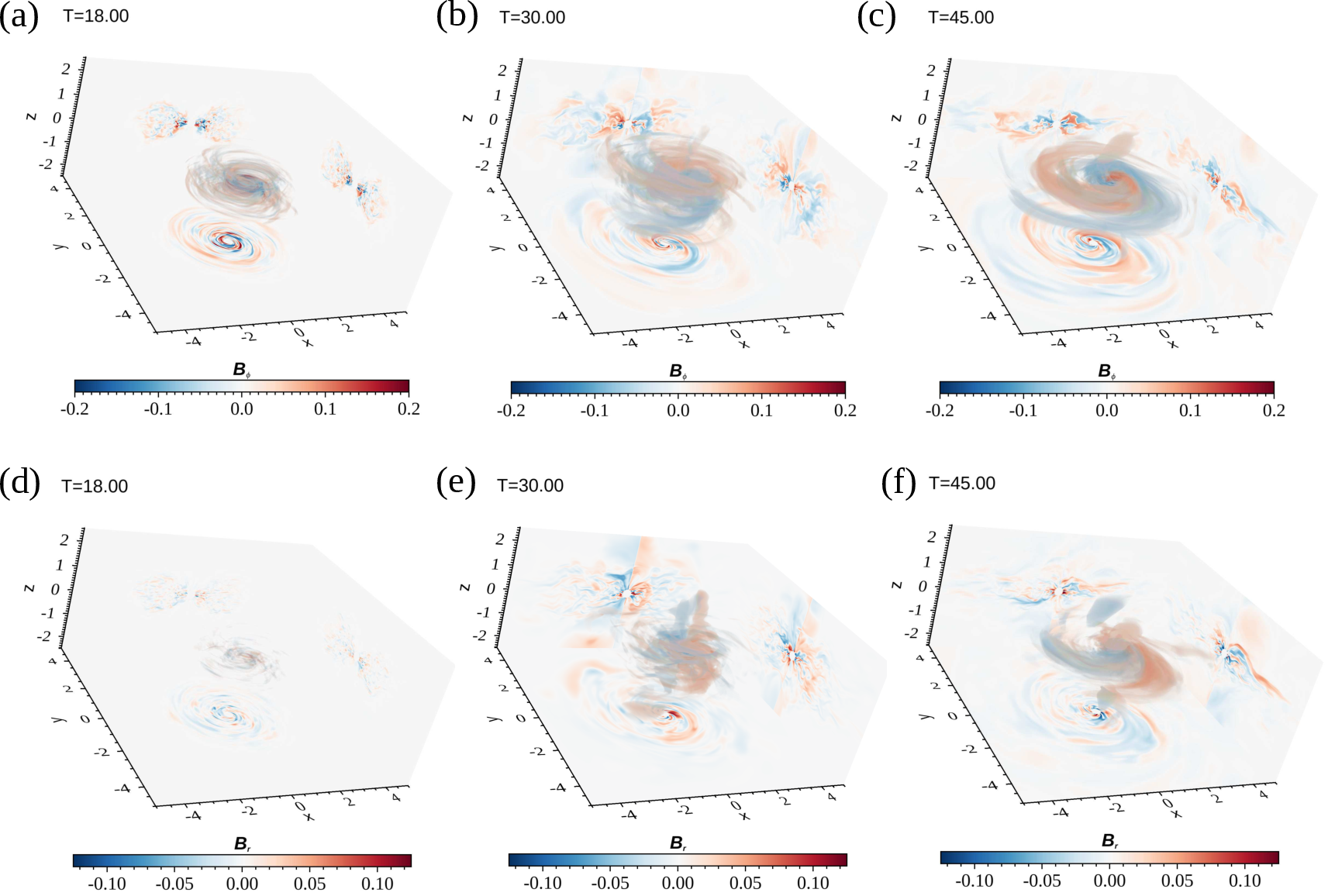}
  \end{center}
  \caption{Time evolution of the accretion disk obtained by the MUSCL scheme for the azimuthal ((a)--(c)) and the radial ((d)--(f)) components of the magnetic field. The format is the same as figures \ref{fig:gMRI_MP5}(d)--\ref{fig:gMRI_MP5}(i)}.
  \label{fig:gMRI_MSCL2}
\end{figure}

When we examined the same problem with the MUSCL scheme, such a dynamo process occurred in the late phase of the evolution, as shown in figure \ref{fig:gMRI_Butterfly}(b). The TVD property of the MUSCL scheme evidently inhibited linear and early nonlinear growths of the MRI. This drawback is also visually recognized in figure \ref{fig:gMRI_MSCL2}, in which magnetic turbulence suffered from strong numerical damping in the early stage (figures \ref{fig:gMRI_MSCL2}(a) and \ref{fig:gMRI_MSCL2}(d)), and global-scale magnetic field survived in the late nonlinear stage (figures \ref{fig:gMRI_MSCL2}(c) and \ref{fig:gMRI_MSCL2}(f)).
\begin{figure}
  \begin{center}
  \includegraphics[scale=0.5]{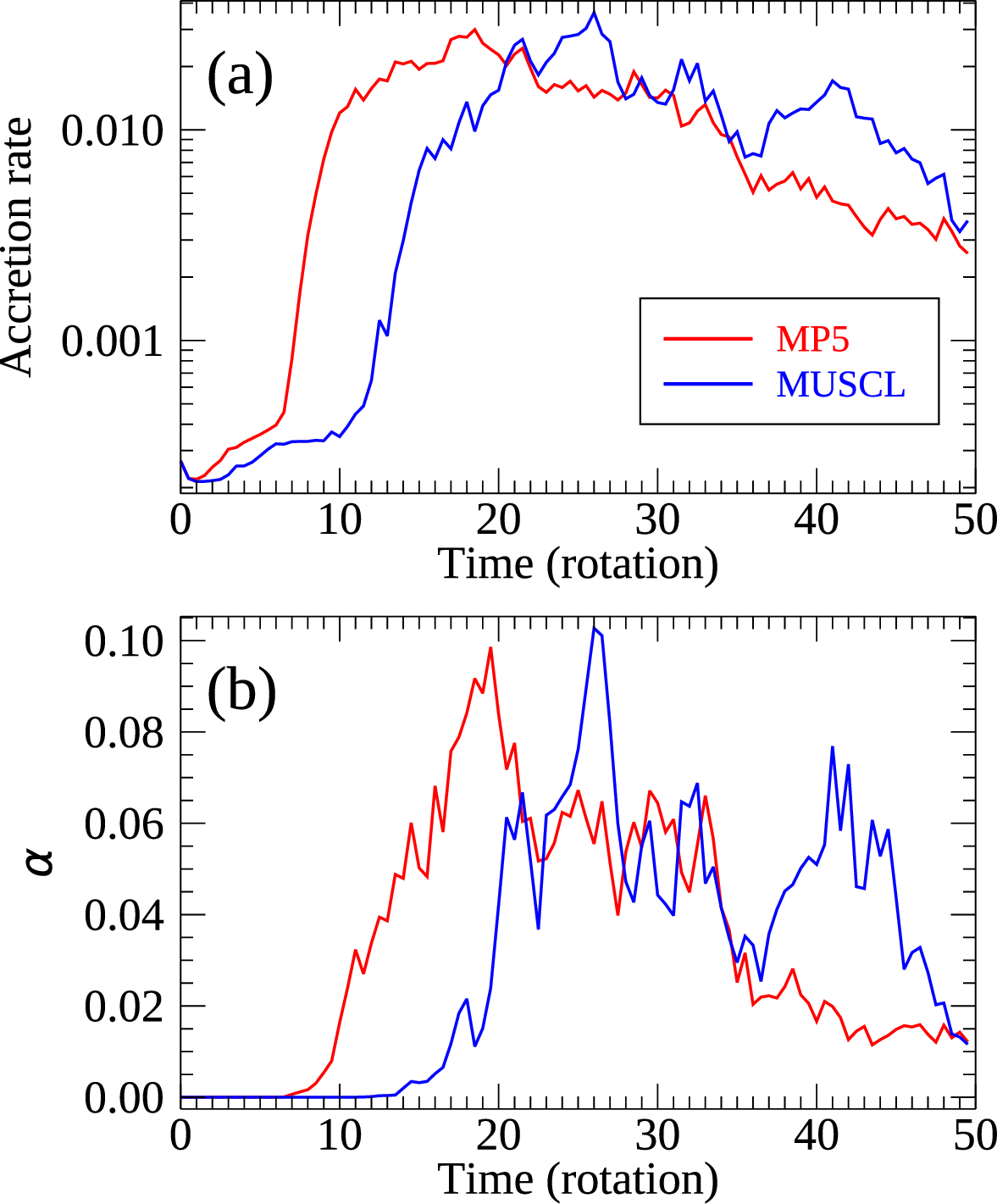}
  \end{center}
  \caption{Time evolution of (a) the absolute value of the the mass accretion rate (eq. \ref{eq:accretion_rate}) and (b) $\alpha$-parameter (eq. \ref{eq:alpha}) obtained by the MP5 (red line) and MUSCL (blue line) schemes.}
  \label{fig:accretion}
\end{figure}																																																																																																						

Figure \ref{fig:accretion}(a) compares the time histories of the mass accretion rate calculated at the inner spherical boundary \citep[cf.][]{Stone2001}

\begin{equation}
\dot{M} = 2\pi r_0^2 \int_0^{\pi} \rho v_r \sin(\theta) d\theta,
\label{eq:accretion_rate}
\end{equation}
where $r_0=0.2$ and $\theta$ are the radius at the inner boundary and the elevation angle in spherical coordinates, respectively. The mass accretion rate peaked at $t\sim19$ and gradually decreased and maintained a certain level until $t=50$ in the run with the MP5 scheme (red line). The mass accretion from the MUSCL scheme (blue line) followed a similar profile but peaked in the late stage of the simulation run because of slow growth of the MRI. 

The mass accretion variations in time coincide with activities of the magnetic dynamo and turbulence in the disk, which can be quantified by the																																																																																																																																		 so-called $\alpha$-parameter \citep{Shakura1973}
\begin{equation}
\alpha = -\frac{\left<B_RB_\phi\right>}{\left<p\right>},
\label{eq:alpha}
\end{equation}
where $<>$ stands for the volume-weighted average of the quantity in cylindrical coordinates. Figure \ref{fig:accretion}(b) compares time histories of the $\alpha$-parameter averaged in a region, $0.2\le R \le 0.5$, $0\le\phi\le 2\pi$, and $-0.25\pi \le \arcsin(z/\sqrt{r^2+z^2}) \le +0.25\pi$. As shown in the mass accretion variation, the $\alpha$-parameter was sustaind at a level of $\alpha \sim 0.015$ following an initial peak of $\alpha \sim 0.1$ at $t=19$ in the MP5 run (red line). The $\alpha$ parameter variation from the MUSCL run (blue line) also exhibited a similar profile as in the MP5 run but with a peak in the late stage as was found in the mass accretion rate variation.

\subsubsection{Convergece test}
\begin{table}
  \caption{Numerical parameters used for numerical convergence tests.}
  \begin{tabular}{|c|c|c|c|c|c|} 
\hline
     & number of cells & $\Delta R$ & $\Delta z$ & range in $R$ & range in $z$ \\ 
\hline
low resolution & $256\times 64 \times 256$ & $1.5\times 10^{-2}$ & $1.5 \times 10^{-2}$ & $0 \le R \le 7.5$ & $-5.8 \le z \le 5.8$ \\
\hline
medium resolution (reference run)& $512\times 128 \times 512$ & $7.5\times 10^{-3}$ & $7.5 \times 10^{-3}$ & $0 \le R \le 9.4$ & $-8.0 \le z \le 8.0$ \\
\hline
high resolution & $768\times 192 \times 768$ & $5.0\times 10^{-3}$ & $5.0 \times 10^{-3}$ & $0 \le R \le 10.1$ & $-8.8 \le z \le 8.8$ \\
\hline
  \end{tabular}
  \label{tab:resolution}
\end{table}

We have examined with the MP5 scheme under different spatial resolutions to assess the numerical convergence property for the present particular problem. The numerical parameters are summarized in Table \ref{tab:resolution} for each spatial resolution.

\begin{figure}
  \begin{center}
  \includegraphics[scale=0.4]{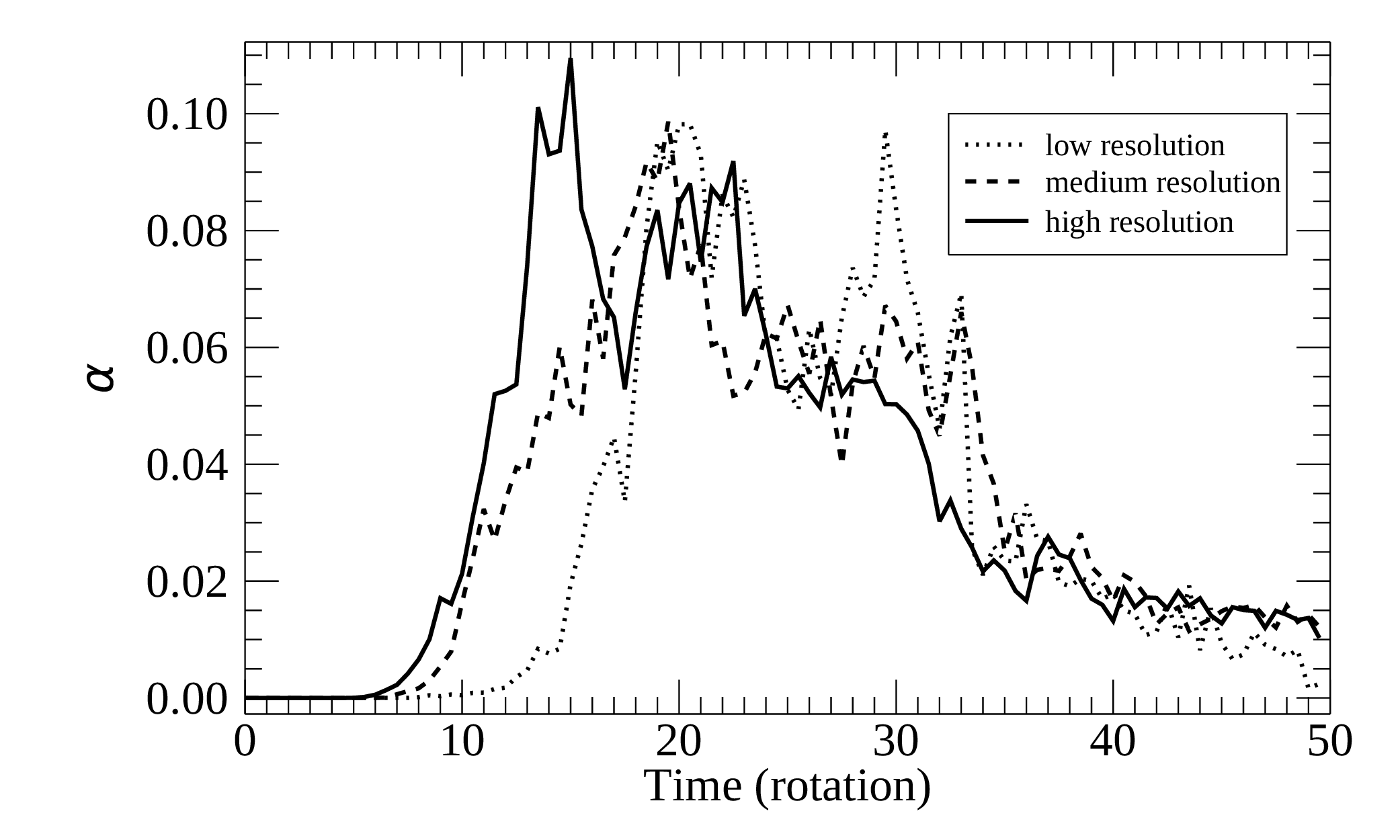}
  \end{center}
  \caption{Time evolution of $\alpha$-parameter (eq. \ref{eq:alpha}) obtained by the MP5 runs with different spatial resolutions. Dotted, dashed, and solid lines represent low-, medium-, and high-resolution runs, respectively. (See also Table \ref{tab:resolution}.)}
  \label{fig:alpha_resolution}
\end{figure}

Figure \ref{fig:alpha_resolution} shows the time evolution of the $\alpha$-parameter for runs with different resolutions in the same format as figure \ref{fig:accretion}(b). Whereas overall the profiles looked similar among runs with different resolutions, the initial growth of the MRI is slow in the low-resolution run (dotted line), as was found in the run with the MUSCL scheme (blue line in figure \ref{fig:accretion}(b)). We could obtain a convergent result of the MRI growth in the early stage up to $t\sim10$ with medium- (dashed line) and high- (solid line) resolution runs. Nevertheless, the spatial resolution used in the high-resolution run is not high enough to give convergent results for further long-term evolution in $t>10$.

\section{Summary and Discussion}
We have developed CANS+, a high-resolution, numerically robust MHD simulation code by employing the HLLD approximate Riemann solver, the MP5 reconstruction method, and the hyperbolic divergence cleaning method. We performed a number of benchmark tests to show the code's capability for solving discontinuities, shock waves, and turbulence all of which are essentially important in astrophysical situations. 

In 1D benchmark tests that included linear Alfv\'en wave propagation and shock tube problems, the adoption of a spatial fifth-order scheme gave superior results compared with a second-order scheme, even when the additional computational costs arising from the higher-order reconstruction were considered: The computation time increased by two times compared with the second-order scheme, but for the same grid resolution the numerical errors from the fifth-order scheme were smaller by orders of magnitude. In other words, to obtain solutions with the same accuracy, the fifth-order scheme required smaller numerical costs by orders of magnitude than the second-order scheme. This advantage is more prominent in multidimensional problems.

In 2D tests of the oblique Alfv\'en wave propagation, the K--H turbulence, the Orszag--Tang vortex problem, and the magnetic reconnection, it was shown that CANS+ enables solving discontinuities, shock waves, and turbulence with high accuracy and stability simultaneously. The test problem of the Parker instability also showed a high capability for solving very low-$\beta$ ($\sim 10^{-3}$) plasma in which the numerical divergence errors of the magnetic field were maintained within reasonably low levels.

As an application of CANS+, we presented global simulations of an accretion disk around a black hole. With a given initial setup and grid resolution, CANS+ was capable of following the long-term evolution of the accretion disk in which the MRI and the resulting mass accretion into the black hole were sustained for $50$ rotational periods. By increasing spatial resolution, we could obtain a convergent result of the early nonlinear growth of the MRI. The low-resolution run with the MP5 scheme gave a similar result to that in the medium-resolution run with the MUSCL scheme. Again, in practice, the MP5 scheme has at least twice the resolution of the MUSCL scheme, giving more than eight times gain in computation time (more than $2^4$ times speed up with doubled computational costs) to obtain results with the same accuracy.

Lastly, we address the caveat of using fifth-order numerical schemes. The characteristic variables used for the reconstruction step showed the best performance with nonoscillatory results in the 1D shock tube problem. With other parameters, such as primitive variables, the updated conservative variables profiles were subject to numerical oscillations around discontinuities even if the MP5 reconstruction gave nonoscillatory profiles at cell surfaces. Thus, the reconstruction step not only required high computational costs because of the variable conversion, but also introduced difficulty in analytically obtaining the eigenvectors and the corresponding eigenvalues of the system. This can be problematic when extending the present MHD to, for example, the special relativistic MHD equations, in which the second-order TVD schemes have been adopted \citep{Matsumoto2011,Takahashi2013}. The search for variables that are universally applicable to high-order reconstruction in various systems of equations remains a task for further applications of CANS+.

\bigskip

\begin{ack}
CANS+ was based on the code originally developed by T. Ogawa at Chiba University. The numerical setup for magnetic reconnection was developed by N. Iwamoto, A. D. Kawamura, J. Sakamoto, and T. Shibayama during the simulation summer school held at Chiba University in 2014. The present simulations used computational resources provided by the Information Technology Center, the University of Tokyo, Research Institute for Information Technology, Kyushu University, and the RIKEN Center for Computational Science through the HPCI System Research project (Project ID: hp120193,  hp120287, hp130027, hp140213, hp140170, hp150263), and Cray XC50 at Center for Computational Astrophysics, National Astronomical Observatory of Japan. This work was supported in part by MEXT SPIRE, MEXT as “Priority Issue on Post-K computer” (Elucidation of the Fundamental Laws and Evolution of the Universe), JICFuS, Research Institute of Stellar Explosive Phenomena at Fukuoka University (JM), JSPS KAKENHI Grant Number 16H03954 (RM) and 17K14260 (HRT).
\end{ack}

\appendix
\section{MP5 reconstruction in nonuniformly spaced cells}
\label{appendix_crt}
A piecewise fourth-degree polynomial is used in the MP5 reconstruction. For the uniformly spaced cells, the left state of a quantity $f$ at a cell $i+1/2$ is, for example, given by
\begin{equation}
^L f_{i+1/2} = \frac{2\bar f_{i-2}-13\bar f_{i-1}+47\bar f_i+27\bar f_{i+1}-3\bar f_{i+2}}{60} = \sum_{r=0}^{4} C_r \bar f _{(i-2+r)},
\label{eq:mp5cc}
\end{equation}
where
\begin{eqnarray}
\bar f_i = \frac{1}{\Delta x_i} \int_{x_{i-1/2}}^{x_{i+1/2}} f(x) dx, \\
\Delta x_{i} = x_{i+1/2}-x_{i-1/2}.
\end{eqnarray}
For more practical uses, in which the cell size is not necessarily uniform, CANS+ employs the Lagrange polynomial as
\begin{equation}
\label{eq:lagrange}
F(x) = \sum_{k=0}^5 \left(\prod_{l=0,l \neq k}^5 \frac{x-x_{i-5/2+l}}{x_{i-5/2+k}-x_{i-5/2+l}}\sum_{m=0}^{k-1} \bar f_{i-2+m}\Delta x_{i-2+m} \right)
\end{equation}
to represent the spatial integral of the quantity $f$ 
\begin{equation}
\label{eq:cumulative_integral}
F(x) = \int_{x_{i-5/2}}^x f(x') dx'.
\end{equation}
Thus, $f$ can be obtained by taking the derivative of equation (\ref{eq:lagrange})
\begin{eqnarray}
f(x) & = & \frac{dF(x)}{dx} \nonumber \\
     & = & \sum_{k=0}^5 \left[ \prod_{l=0,l \ne k}^5 \frac{x-x_{i-5/2+l}}{x_{i-5/2+k}-x_{i-5/2+l}} \sum_{m=0,m \ne k}^5 \frac{1}{x-x_{i-5/2+m}} \sum_{n=0}^{k-1} \bar f_{i-2+n}\Delta x_{i-2+n}\right].
\end{eqnarray}
Finally, the coefficients $C_r$ in equation (\ref{eq:mp5cc}) can be obtained for $x_{i+1/2}$ from $f(x_{i+1/2})$ as
\begin{equation}
C_r = \sum_{k=1+r}^{5} \left[ \prod_{l=0,l \ne k}^5 \frac{x_{i+1/2}-x_{i-5/2+l}}{x_{i-5/2+k}-x_{i-5/2+l}} \sum_{m=0,m \ne k}^5 \frac{1}{x_{i+1/2}-x_{i-5/2+m}} \Delta x_{i-3+k}\right].
\label{eq:mp5_orig}
\end{equation}
Note that $C_r$ depends only on the cell size $\Delta x_{i+n}$ and the cell surface's location $x_{i+n/2}$. Thus, they can be determined at initialization.

\subsection*{Benchmark test}

\begin{figure}
  \begin{center}
  \includegraphics[scale=0.8]{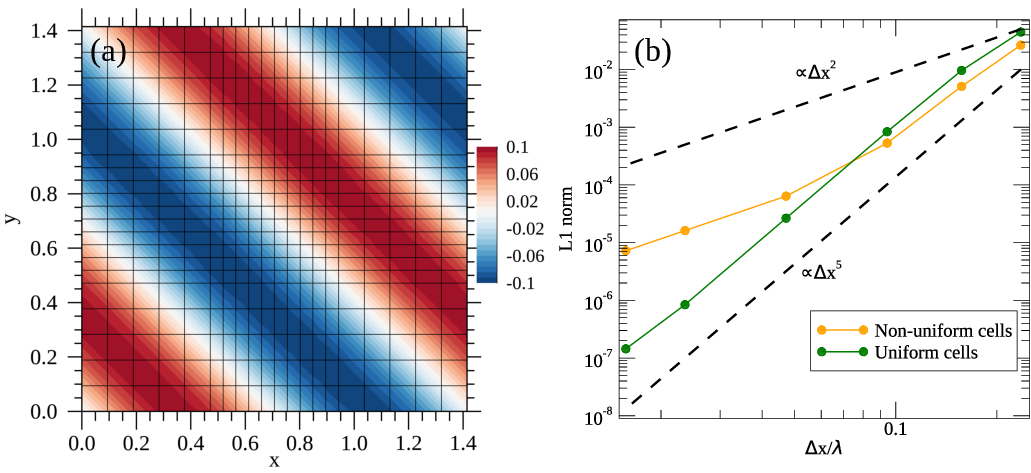}
  \end{center}
  \caption{Circularly polarized linear Alfv\'en wave propagation tests in the oblique (diagonal) propagation in nonuniformly spaced cells. (a) $B_z$ profile after five Alfv\'en transit times along with the cells' shape. (b) $L_1$ norm errors obtained from numerical experiments with different cell resolutions ($\Delta h_c$) for the uniform (green) and nonuniform (orange) cases. The dashed lines indicate the order of accuracy in space.}
  \label{fig:2DCPA_nonuniform}
\end{figure}

We examined a circularly polarized Alfv\'en wave propagation in two dimensions using nonuniformly spaced cells as a benchmark test of the interpolation procedure of equation (\ref{eq:mp5_orig}). The initial setup is the same as that presented in sub-subsection \ref{2DCPA}, except we adopted the propagation angle of $\theta=45\degree$. In this test, we used the following two different cell sizes
\begin{equation}
\Delta x = \Delta y = \Bigg\{
\begin{array}{l}
\Delta h_c \quad (0 \le x, y \le \frac{L_{x,y}}{3}, \frac{2}{3}L_{x,y} < x, y \le L_{x,y} )\\
\frac{\Delta h_c}{2} \quad (\frac{L_{x,y}}{3} < x, y \le \frac{2}{3}L_{x,y}) 
\end{array},
\end{equation}
where $L_{x,y}$ is the system's size in the $x$- and $y$-directions.

By examining with different sizes of $\Delta h_c$, the accuracy of the code in nonuniformly spaced cells was obtained. We calculated the $L_1$ norm error by the following equation
\begin{equation}
L_1\ {\rm norm} = \frac{1}{L_xL_y} \sum_{i,j} \Delta_{x,i}\Delta_{y,j}\left|B_{z(i,j)}-B_{0(i,j)}\right|,
\end{equation}
where $(i,j)$ indicates the cell number in two dimensions and $B_{0(i,j)}$ is the analytic solution at each cell-center location.

Figure \ref{fig:2DCPA_nonuniform}(a) shows the 2D profiles of the $z$-component of the magnetic field after five Alfv\'en transit times along with the cells' shape. We determined that the code can solve smooth profiles in nonuniformly spaced cells without strong damping of the wave amplitude and large phase errors. Figure \ref{fig:2DCPA_nonuniform}(b) shows the $L_1$ norm errors for various $\Delta h_c$ for the nonuniform (orange line) cell cases. Here, we also plotted results from uniform cell cases (green line). We found that the error decreased following the slope expected from the fifth-order interpolation for $\Delta h_c/\lambda > 0.1$. We also note that the errors are smaller in the nonuniform cases than in the uniform cases. However, the accuracy curve approaches the slope of the second-order accuracy for $\Delta h_c/\lambda \le 0.1$ in the nonuniform cases. The accuracy curve for the uniform cases followed the fifth-order slope all the way down to $\Delta h_c/\lambda \sim 0.01$, as expected from the results in sub-subsection \ref{2DCPA}.

\section{MP5 reconstruction in cylindrical coordinates}
\label{appendix_cyl}
\cite{Mignone2014} showed that incorporating the curvature of the cell into the piecewise polynomial reconstruction in curvilinear coordinates, namely, the volume-weighted reconstruction, improved the solutions near the origin of the coordinate axis (along the $z$-axis in cylindrical coordinates). This curvature effect is considered for the MP5 reconstruction in the $R$-direction in cylindrical coordinates. In this case, equations (\ref{eq:lagrange}) and (\ref{eq:cumulative_integral}) are modified as
\begin{eqnarray}
F(R) &=& \int_{R_{i-5/2}}^R R'f(R') dR'\\ \nonumber
    &=& \sum_{k=0}^5 \left(\prod_{l=0,l \neq k}^5 \frac{R-R_{i-5/2+l}}{R_{i-5/2+k}-R_{i-5/2+l}}\sum_{m=0}^{k-1} \bar f_{i-2+m}R_{i-2+m}\Delta R_{i-2+m} \right) ,
\end{eqnarray}
where
\begin{equation}
\bar f_i = \frac{1}{R_i\Delta R_i} \int_{R_{i-1/2}}^{R_{i+1/2}} Rf(R) dR
\end{equation}
\begin{equation}
R_i = \frac{R_{i+1/2}+R_{i-1/2}}{2},
\label{eq:cyl_volume}
\end{equation}
\begin{equation}
\Delta R_{i} = R_{i+1/2}-R_{i-1/2}. 
\label{eq:cyl_volume2}
\end{equation}
As for Cartesian coordinates, the coefficients $C_r$ are similarly obtained for $R_{i+1/2}$ from $f(R_{i+1/2})=\frac{1}{R_{i+1/2}}\frac{dF}{dR}|_{R=R_{i+1/2}}$ as
\begin{equation}
C_r = \frac{1}{R_{i+1/2}}\sum_{k=1+r}^{5} \left[ \prod_{l=0,l \ne k}^5 \frac{R_{i+1/2}-R_{i-5/2+l}}{R_{i-5/2+k}-R_{i-5/2+l}} \sum_{m=0,m \ne k}^5 \frac{1}{R_{i+1/2}-R_{i-5/2+m}} R_{i-3+k}\Delta R_{i-3+k}\right].
\label{eq:mp5_cyl}
\end{equation}
Note that the coefficient for the right state of the cell surface at $R=0$ becomes infinity. This singularity at $R=0$ is addressed by adopting the first-order reconstruction, i.e., $^R f_0 = \bar f_1$. 

\subsection*{Boundary condition at $R=0$}
In the following, we present results from benchmark tests to discuss how the two interpolation procedures of equations (\ref{eq:mp5_orig}) and (\ref{eq:mp5_cyl}) result in different evolutions in cylindrical coordinates. 

For boundary conditions across the $z$-axis, $\rho$, $p$, and ($V_z,\ B_z$) were assumed axisymmetric profiles, whereas the antisymmetric boundary condition is applied to ($V_R,\ V_\phi,\ B_R,\ B_\phi$) for the case with equation (\ref{eq:mp5_orig}) as usual. Conversely, with equation (\ref{eq:mp5_cyl}), the symmetric and antisymmetric boundary conditions were applied to ($V_R,\ V_\phi,\ B_R,\ B_\phi$) and ($V_z$,\ $B_z$), respectively. We found this somewhat odd boundary condition resulted a best practice with equation (\ref{eq:mp5_cyl}) in the following experiments. A naive idea behind this is that a cell volume $R_i\Delta R_i$ becomes negative in $R<0$ by definition (equations (\ref{eq:cyl_volume}) and (\ref{eq:cyl_volume2})), and the sign of the vector quantities should be reversed when constructing the volume-weighted polynomial. Note, however, that the usual boundary condition should be used for other situations, such as when calculating the current density (Appendix \ref{diffusion_test}). The special care for the boundary condition in cylindrical coordinates was also addressed in global simulations of a black hole accretion disk in Section \ref{gMRI}.

\subsection*{Benchmark tests}
\subsubsection*{1D magnetic confinement of a cylindrical plasma column}
\begin{figure}
  \begin{center}
  \includegraphics[scale=0.6]{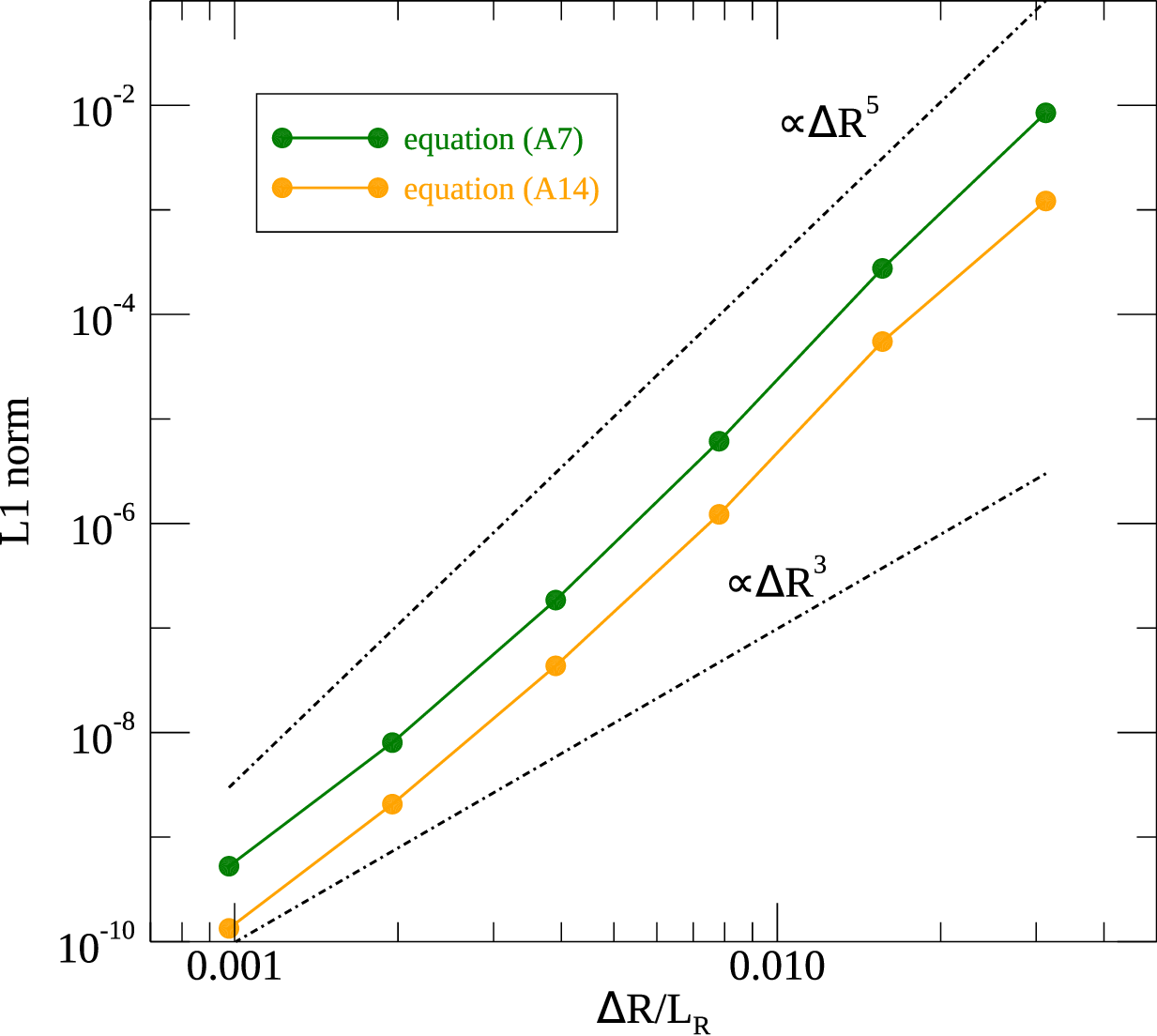}
  \end{center}
  \caption{1D magnetic confinement of a cylindrical plasma column was examined with various cell sizes. $L_1$ norm errors are shown for the normal (equation (\ref{eq:mp5_orig}), green) and volume-weighted (equation (\ref{eq:mp5_cyl}), orange) MP5 reconstructions. Dashed lines indicate the orders of accuracy in space.}
  \label{fig:mag_confine}
\end{figure}

First, we examined a static balance problem presented by \citet{Mignone2014}. In this test, a cylindrical plasma column was initially confined by the toroidal magnetic field of $B_\phi$, satisfying a radial force balance between the gas pressure ($p$) gradient and the Lorentz forces, as shown in the following equations
\begin{eqnarray}
p =  \frac{p_0}{\left(1+R^2/R_0^2\right)^2},\\
B_\phi = \frac{R\sqrt{2p_0}}{R_0\left(1+R^2/R_0^2\right)}, 
\end{eqnarray}
where $p_0=1$ at $R=0$ and $B_\phi=1/\sqrt{2}$ at $R_0=1$. The initial cell-averaged quantities were numerically obtained by Simpson's rule.

The solutions were obtained in a 1D domain in $0 \le R \le L_R$ for different spatial resolutions at a normalized time of $t=10$. Here, the domain size was $L_R=10$, and we used $32$ to $1024$ computational cells to obtain the spatial accuracy of the code. The CFL number was fixed to $\sigma_c = 0.05$.

For the present cylindrical case, the $L_1$ norm error was calculated by
\begin{equation}
L_1\ {\rm norm} = \frac{1}{\Delta_{cyl}} \sum_{i}^{max} R_i\left|V_{R,i}(t=10)-V_{R,i}(t=0)\right|,
\end{equation}
where $\Delta_{cyl}$ is the cylindrical volume of a 1D domain $0\le R_i \le R_{max}$:
\begin{equation}
\Delta_{cyl} = \int_0^{R_{max}} R dR = \frac{R_{max}^2}{2},
\end{equation}
and $i$ indicates the cell number. To highlight the differences between the two schemes, we adopted $R_{max}=1.0$. Obviously, $V_R(t=0)=0$ in the present static balance problem.

Figure \ref{fig:mag_confine} shows the $L_1$ norm errors with different spatial resolutions for the interpolations of equation (\ref{eq:mp5_orig}) (green line) and equation (\ref{eq:mp5_cyl}) (orange line). The overall spatial accuracy follows a slope expected from the fifth-order interpolation, but the accuracy curve approaches the third order of the SSP--RK time integration in high resolution runs ($0.004 < \Delta R/L_R$) under the fixed CFL condition. The error with the interpolation of equation (\ref{eq:mp5_orig}) is, however, about one order of magnitude larger than the error of the interpolation incorporating the cell's curvature effect (equation (\ref{eq:mp5_cyl})) in all spatial resolutions. We found great improvement in the solution, especially near the coordinate origin.

\subsubsection*{2D blast-wave propagation}

\begin{figure}
  \begin{center}
  \includegraphics[scale=0.8]{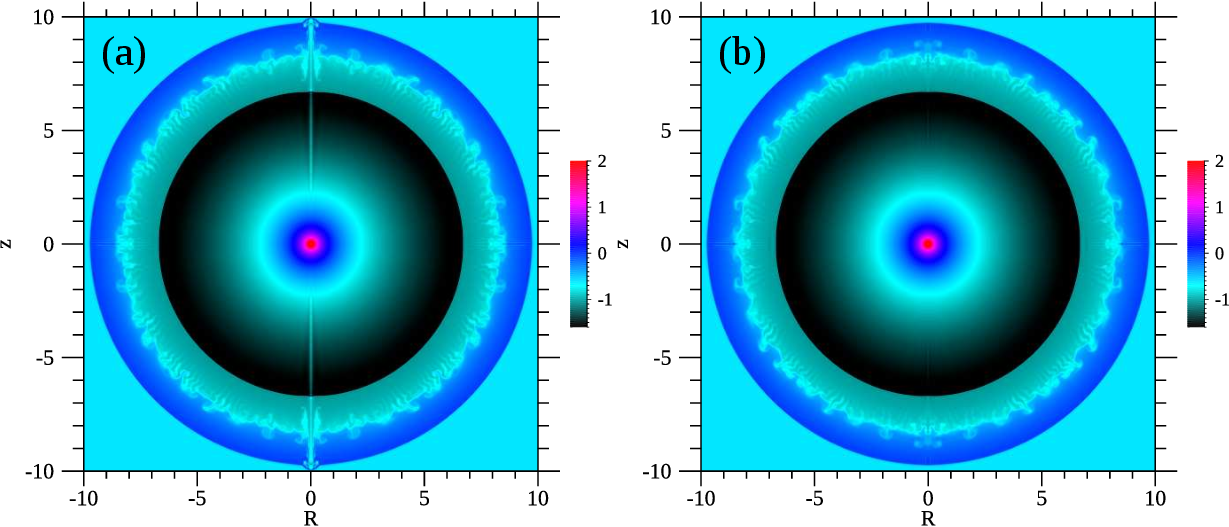}
  \end{center}
  \caption{Blast-wave propagation test in 2D cylindrical coordinates (R, z). Mass density profiles in a logarithmic scale are shown for the reconstruction procedures of (a) equation (\ref{eq:mp5_orig}) and (b) equation (\ref{eq:mp5_cyl}).}
  \label{fig:wind}
\end{figure}

Next we demonstrate a blast-wave propagation in a 2D ($R$--$z$) configuration. The initial setup is the same as that introduced by \citet{Mignone2014}, and constant background quantities were given as 
\begin{eqnarray}
\rho_0 &=& \frac{1}{4}, \\
V_R &=& V_z = 0,  \\
p&=&C_{s0}^2\frac{\rho}{\gamma},
\end{eqnarray}
where $C_{s0}=4\times 10^{-3}$ is the sound speed in the ambient medium. Inside the spherical region of $0\le \sqrt{R^2+z^2} \le 1.0$, 
\begin{eqnarray}
\rho V_R R^2 &=& 1.0 \\
V_R &=& \tanh\left(\frac{\sqrt{R^2+z^2}}{0.2}\right)\frac{R}{\sqrt{R^2+z^2}}, \\
V_z &=& \tanh\left(\frac{\sqrt{R^2+z^2}}{0.2}\right)\frac{z}{\sqrt{R^2+z^2}}, \\
p &=& \frac{\rho^\gamma C_{s1}^2}{\gamma},
\end{eqnarray}
where $C_{s1}=3\times 10^{-2}$ is the sound speed in the wind region. The initial cell-averaged quantities for these configurations were numerically obtained by Simpson's rule. The profiles in the spherical region were set constant in time so  a supersonic flow blows out from the inner spherical region. Initialized by this setup, the time evolution was solved in a simulation domain that covered $0 \le R \le 10$ and $-10 \le z \le 10$  with $256 \times 512$ computational cells ($\Delta R = \Delta z = 1.95\times10^{-2}$). The open boundary condition was applied at $R=10.0$ and $|z|=10.0$. 

When using equation (\ref{eq:mp5_orig}) for the reconstruction in the $R$-direction, the forward shock was subject to a spurious deformation in addition to a trail in the rarefied region along the $z$-axis (figure \ref{fig:wind}(a)). The growth of the numerical artifacts near the $z$-axis ceased and the solution was greatly improved (figure \ref{fig:wind}(b)) by incorporating the curvature effect of the cell into the reconstruction (equation (\ref{eq:mp5_cyl})). Turbulent evolution of the Rayleigh--Taylor instability at the contact discontinuity between the sharp shock wave fronts highlights the superior capability of CANS+, which employs the MP5 scheme and the HLLD approximate Riemann solver.

\section{Accuracy of the resistivity terms}
\label{diffusion_test}
In this section, we present a convergence test of the magnetic diffusion problem presented by \cite{MatsumotoTomoaki2011} to evaluate the source term representation associated with a finite magnetic resistivity (equations (\ref{eq:induction}) and (\ref{eq:energy})).

The current density $\boldsymbol{j}=\nabla \times \boldsymbol{B}$ is evaluated by the second-order central finite difference, for example, 
\begin{equation}
j_{x(i,j)} = \frac{\partial B_z}{\partial y} = \frac{B_{z(i,j+1)}-B_{z(i,j-1)}}{2\Delta y},
\end{equation}
where $\Delta y$ is the cell size in the $y$-direction and $(i,j)$ indicates the cell number in two dimensions. Then, the current density at the cell surface is obtained by the arithmetic average of the two neighboring cell-center values and is added to the numerical flux $F^{*}$ of the ideal MHD part
\begin{equation}
F^{*}_{i,j+1/2} = F^{*}_{i,j+1/2}+\frac{\eta_{(i,j)}j_{x(i,j)}+\eta_{(i,j+1)}j_{x(i,j+1)}}{2}.
\end{equation}

\begin{figure}
  \begin{center}
  \includegraphics[scale=0.4]{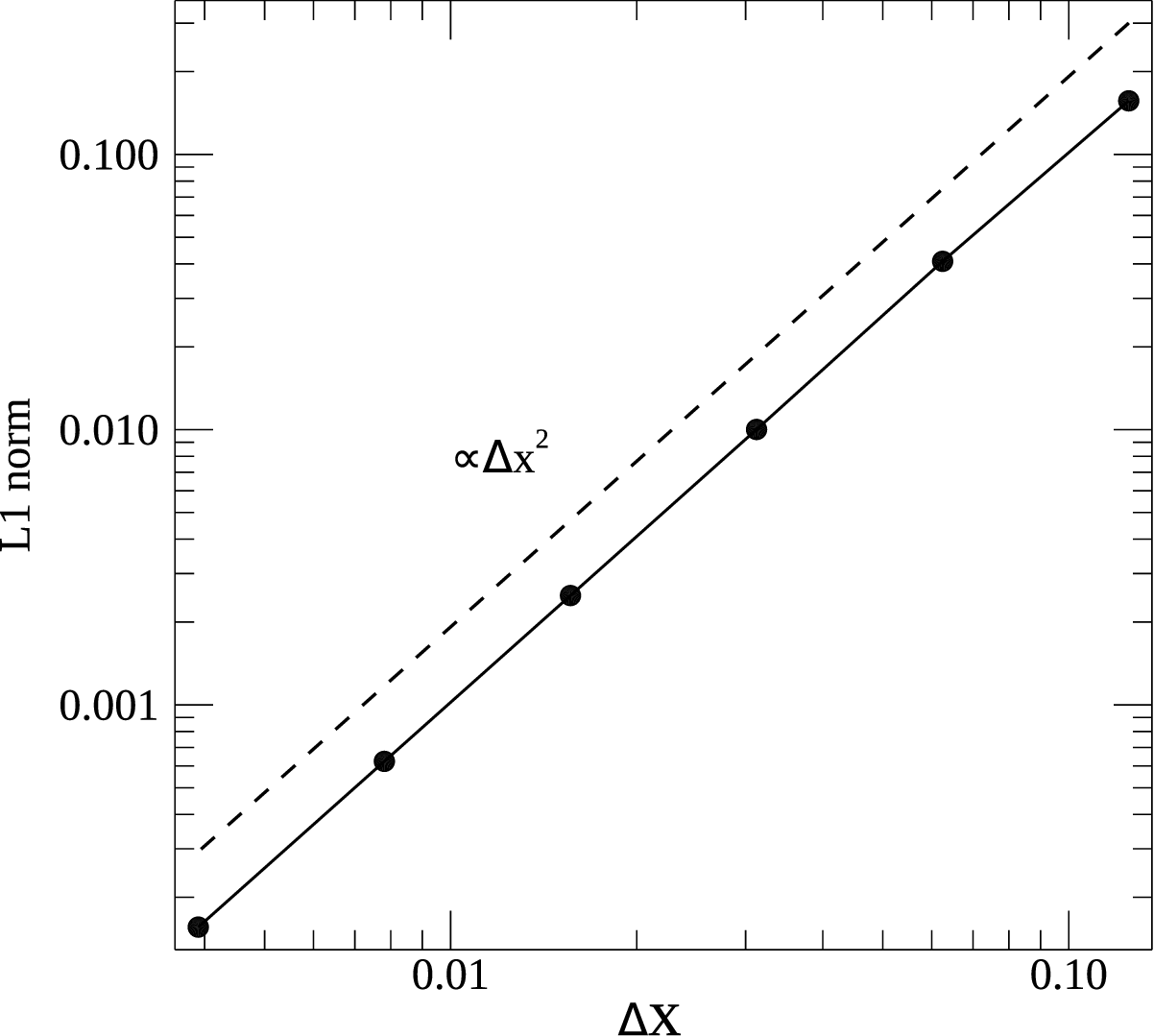}
  \end{center}
  \caption{$L_1$ norm errors of the magnetic diffusion problem for various cell sizes. The dashed line indicates the second-order accuracy in space.}
  \label{fig:diffusion}
\end{figure}

Here, we only focus on the diffusion term in the induction equation ($F^{*}=0$)
\begin{equation}
\frac{\partial \boldsymbol{B}}{\partial t} = \nabla \times \left(\eta \boldsymbol{j}\right) = \nabla \times \left(\eta \nabla \times \boldsymbol{B} \right).
\end{equation}
This equation was solved in a 2D plane in $0\le x\le 1$ and $0\le y\le 0.5$ with the periodic boundary condition in each direction. The $z$-component of the magnetic field was initially provided by
\begin{equation}
B_z = \sin(\boldsymbol{k}\cdot\boldsymbol{r}),
\end{equation}
where $\boldsymbol{k} = 2\pi(1,\ 2)^T$ and $\boldsymbol{r} = (x,\ y)^T$. The analytic solution to the present diffusion problem can be obtained as
\begin{equation}
B_{z}(t) = \exp(-\eta|\boldsymbol{k}^2|t)\sin(\boldsymbol{k}\cdot\boldsymbol{r}),
\end{equation}
with which $L_1$ norm errors were calculated at a time of $t=4\times 10^{-3}$ with various cell sizes from $\Delta x = \Delta y = 1/8$ to $\Delta x = \Delta y = 1/256$ under a fixed diffusion number of $\sigma_d=0.3$. We adopted $\eta=1.0$.

Figure \ref{fig:diffusion} shows the $L_1$ norm errors as a function of various cell sizes. As expected, the errors decreased as the cell size gets smaller, following the second-order accuracy slope. Nevertheless, the resistivity terms become important in localized regions inside current layers (e.g., the magnetic reconnection region shown in sub-subsection \ref{reconnection}), and the impacts of the lower-order representation to overall spatial accuracy are generally limited.



\end{document}